\documentclass[twocolumn]{aastex631}

\usepackage{xspace}
\usepackage{amsmath}
\usepackage{framed} 
\usepackage{txfonts}
\usepackage{epstopdf}
\usepackage{rotating}
\usepackage{natbib}
\usepackage{ulem}
\usepackage{txfonts}
\usepackage{color}
\usepackage{longtable}
\usepackage{float}

\renewcommand{\edit}[1]{{\textcolor{black}{#1}}}

\shorttitle{Binary Microlensing Populations}
\shortauthors{Abrams et al.}

\begin{document}

\title{Assessing the Impact of Binary Systems on Microlensing Using \texttt{SPISEA} and \texttt{PopSyCLE} Population Simulations}
\author[0000-0002-0287-3783]{Natasha~S.~Abrams}
\correspondingauthor{Natasha~S.~Abrams}
\email{nsabrams@berkeley.edu}
\affiliation{University of California, Berkeley, Department of Astronomy, Berkeley, CA 94720}

\author[0000-0001-9611-0009]{Jessica~R.~Lu}
\affiliation{University of California, Berkeley, Department of Astronomy, Berkeley, CA 94720}

\author[0000-0002-6406-1924]{Casey~Y.~Lam}
\affiliation{University of California, Berkeley, Department of Astronomy, Berkeley, CA 94720}
\affiliation{Observatories of the Carnegie Institution for Science, Pasadena, CA 91101, USA}

\author[0000-0002-7226-0659]{Michael~S.~Medford}
\affiliation{University of California, Berkeley, Department of Astronomy, Berkeley, CA 94720}

\author[0000-0003-2874-1196]{Matthew~W.~Hosek,~Jr.}
\affiliation{University of California, Los Angeles, Department of Astronomy, Los Angeles, CA 90095}

\author[0000-0003-4725-4481]{Sam~Rose}
\affiliation{University of California, Berkeley, Department of Astronomy, Berkeley, CA 94720}
\affiliation{Cahill Center for Astronomy and Astrophysics, California Institute of Technology, Pasadena, CA 91125, USA}

\date{\today}

\begin{abstract}
Gravitational microlensing provides a unique opportunity to probe the mass distribution of stars, black holes, and other objects in the Milky Way. Population simulations are necessary to interpret results from microlensing surveys. The contribution from binary objects is often neglected or minimized in analysis of observations and simulations despite the high percentage of binary systems and microlensing's ability to probe binaries. To simulate the population effects we added multiple systems 
to Stellar Population Interface for Stellar Evolution and Atmospheres (\texttt{SPISEA}), \edit{which} simulates stellar clusters. We then inject these multiples into Population Synthesis for Compact-object Lensing Events (\texttt{PopSyCLE}), \edit{which} simulates Milky Way microlensing surveys. When making OGLE observational selection criteria, we find that 55\% of observed microlensing events involve a binary system. Specifically, 14.5\% of events have a multiple-lens and single source, 31.7\% have a single lens and a multiple-source, and 8.8\% have a multiple-lens and a multiple-source. The majority of these events have photometric lightcurves that appear single \edit{and are fit well by a single-lens, single-source model}. This suggests that binary source and binary lens-binary source models should be included more frequently 
in event analysis. 
The mean Einstein crossing time shifts from 19.1 days for single events only to 21.3 days for singles and multiple events, after cutting binary events with multiple peaks. The Einstein crossing time distribution of singles and single-peaked multiple events is better aligned with observed distributions from OGLE \citep{2017Natur.548..183M} than singles alone, indicating that multiple systems are a significant missing piece between simulations and reality.
\end{abstract}


\section{Introduction}
\label{sec:introduction}
Gravitational lensing is when a mass, such as a black hole or star, acts as a lens and bends the light of a background star, magnifying it and creating multiple images. In microlensing we are not able to resolve the individual images, so instead we can see a characteristic brightening and dimming of the source star \citep{1986ApJ...304....1P, 1991ApJ...372L..79G}. Gravitational microlensing is a powerful phenomenon which we can used for a variety of applications. It can be used to answer Galactic structure questions \citep{Rich:2013} such as the existence of a Galactic bar \citep{KiragaPacynski:1994, Hamadache:2006, Moniez:2017} and whether the Milky Way's potential is triaxial \citep{Zhao:1996}. It can constrain the population of objects in the Galactic bulge such as dwarf stars which act as lensed sources \citep{Bensby:2013} and the Initial Mass Function of the Bulge \citep{Wegg:2017}. It also has the ability to probe the mass distribution of dim objects in the Milky Way, such as exoplanets \citep[e.g.][]{1998ApJ...500...37G, 2008ApJ...684..663B, Gaudi:2012, Penny:2019}, brown dwarfs \citep[e.g.][]{Park:2013, Street:2013}, and black holes \citep[e.g.][]{2002astro.ph..7006B, 2002MNRAS.329..349M, 2016ApJ...830...41L, 2018ApJ...867...37K, 2019IAUS..339...16W,2019ApJ...885....1W, 2020A&A...636A..20W}. 

For isolated black holes in particular, since microlensing only depends on their gravitational nature and does not require them to interact or be in a binary system, microlensing is the only method to probe them. There is one isolated black hole candidate that has been identified via microlensing \citep{Lam:2022, Sahu:2022, Mroz:2022, Lam:2023-OB110462}. The mass of only a few dozen other black holes have been measured, all of which were in binaries. These include X-ray binaries \citep{2016A&A...587A..61C}, merging black holes whose gravitational waves were detected by the Laser Interferometer Gravitational-Wave Observatory (LIGO) \citep{PhysRevLett.116.061102}, and non-interacting black holes in binaries found via their astrometric signals \citep{El-Badry:2023-BH1, El-Badry:2023-BH2, Chakrabarti:2023}. To interpret these findings and put them in a broader astrophysical context, we need to understand the mass distribution, binary fraction, and velocity distributions for all black holes. Uncovering the underlying distribution of black holes in the Milky Way is important to understand the process of stellar evolution, supernovae, potential seeds for super massive black holes at the center of galaxies, and could even give insight into the nature of dark matter. 

Numerous photometric surveys have been used to find microlensing events.
Historically, the MACHO \citep{2000ApJ...542..281A} and EROS \citep{2007A&A...469..387T} projects were used to search for dark matter candidates in the Galactic halo through microlensing.
More recently, surveys mainly targeting the Galactic bulge are primarily focused on exoplanet microlensing (i.e. OGLE \citep{ogleIV:Udalski:2015}, MOA \citep{moa:Sumi:2003}, KMTNet \citep{kmtnet:kim:2018}). There are also other surveys which are not exclusively focused on microlensing which have been used to discover microlensing events (i.e. Gaia \citep{Wyrzykowski:2023}, ZTF \citep{Rodriguez:2022, Medford:2023, Zhai:2023}, and VVV \citep{Navarro:2020, Husseiniova:2021, Kaczmarek:2022, Kaczmarek:2023}). Combined, these surveys find thousands of events each year. 

When interpreting these microlensing events, multiple systems have been neglected except when they produce an obvious signal; however, most stars are in multiple star systems \citep[e.g.][]{2013ARA&A..51..269D}. There are some physical phenomena which require the presence of multiple systems, including the production of hypervelocity stars \citep{Rasskazov:2019}, and some which require including multiple systems in our interpretations, such as the occurrence rates of exoplanet observations \citep{MoeKratter:2021}. As such, it is critical to include them when interpreting microlensing surveys, both on the individual event level and population level. As we explore in this paper, the inclusion of multiples can significantly affect the inferred event rates and distributions of microlensing event parameters.

The rest of the paper is organized as follows. In Section \ref{sec:methods}, we explain how we add binary and triple systems into our microlensing simulation. In Section \ref{sec: computation}, we detail how the simulations were run and what observational cuts we make to compare them to data. In Section \ref{sec:results}, we discuss our results including the fraction of microlensing events we find, how the Einstein crossing time distribution and event rates compare to observational data, the properties of multiple systems before and after microlensing, and how adding multiple systems affects black hole astrometric candidate selection. In Section \ref{sec:discussion}, we compare our analysis to other simulations and discuss further improvements. In Section \ref{sec:conclusion}, we summarize our conclusions.

\section{Adding Multiple Systems to \texttt{SPISEA} and \texttt{PopSyCLE}}
\label{sec:methods}
In Section \ref{sec: binary params}, we describe the parameters of binaries input into Stellar Population Interface for Stellar Evolution and Atmospheres \citep[\texttt{SPISEA},][]{2020AJ....160..143H}, a population synthesis code used to generate single-age, metallicity clusters. The companion objects are then injected into Population Synthesis for Compact-object Lensing Events \citep[\texttt{PopSyCLE}, ][]{2020ApJ...889...31L} and we carry out a mock-microlensing survey including these multiple systems, as described in Section \ref{sec: popsycle}. There were a number of additional improvements made to \texttt{PopSyCLE}. See the \texttt{PopSyCLE} change log for more details.\footnote{\url{https://popsycle.readthedocs.io/en/latest/source/changelog.html}}

\subsection{Binary population parameters}
\label{sec: binary params}
\begin{figure}
    \centering
    \includegraphics[width=0.48\textwidth]{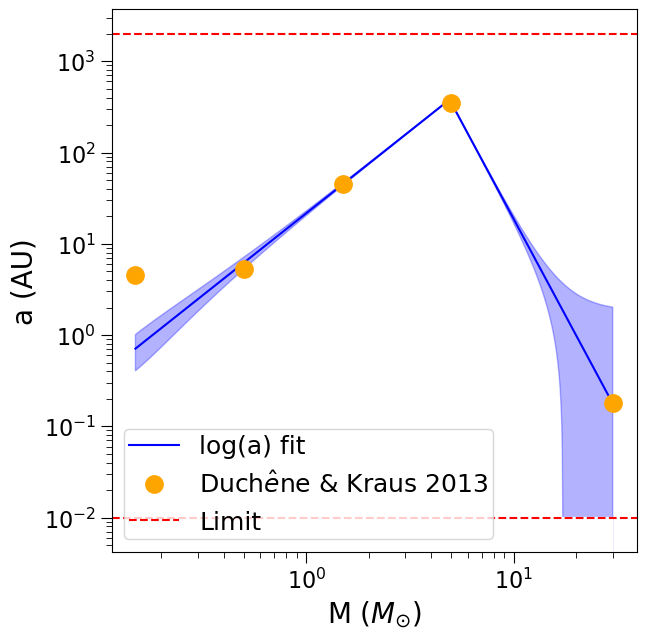}
    \caption{Semi-major axis ($a$)  as a function of primary mass ($M$). The orange points are the mean semi-major axis values averaged over mass bins in \cite{2013ARA&A..51..269D}. We fit these points and corresponding standard deviations, shown in blue. At each mass, the semi-major axis distribution is modeled as a log-normal distribution with a mean (indicated by the dark blue line) and standard deviation (indicated by the shaded blue region). They were fit as a broken power law described in Equation \ref{broken_PL}. We set a lower limit of $10^{-2}$ AU \cite[see Fig.~2 of ][]{2013ARA&A..51..269D} and an upper limit of 2000 AU \cite[see Table 2 of ][]{2013ARA&A..51..269D}.}
    \label{fig:a}
\end{figure}

\texttt{SPISEA} is a population synthesis code that generates a star cluster with customizable age, metallicity, and stellar evolution, among other characteristics. The initial version of \texttt{SPISEA} \citep{2020AJ....160..143H} only modeled unresolved stellar multiplicity and its effect on synthetic photometry. In this work, we add functionality to model resolved stellar multiplicity by adding semi-major axis, eccentricity, and orbital parameters for individual systems (see Table \ref{tab: spisea parameters}).

\texttt{SPISEA} has a default or tunable mass-dependent multiplicity fraction, companion star frequency, and mass ratio (\citealp{2020AJ....160..143H}; see also \citealp{2013ApJ...764..155L}).
In \texttt{SPISEA}, the primary is drawn from the initial-mass function \edit{(IMF)} and is the object with the largest initial mass in the system and the companions are the other less massive objects in the system.

The multiplicity fraction (MF) is the fraction of systems with multiple stars which is a function of the primary mass ($M$, in solar masses):
\begin{equation}
    \label{MF}
    \textrm{MF}(M) = AM^{\alpha},
\end{equation}
where $A = 0.44$ and $\alpha = 0.51$ by default \citep{2013ApJ...764..155L}.
The companion star fraction (CSF) is the number of stars per multiple system which is also a function of mass:
\begin{equation}
    \label{CSF}
    \textrm{CSF}(M) = BM^{\beta},
\end{equation}
where $B = 0.5$ and $\beta = 0.45$ by default \citep{2013ApJ...764..155L}.
In \texttt{SPISEA} we can set a probabilistic or hard limit on the number of companions per system. 
To assign each companion mass, \texttt{SPISEA} selects from a distribution of mass ratios, $q$, where
\begin{equation}
    q = \frac{M_{\rm companion}}{M_{\rm primary}}.
\end{equation}
The default mass-ratio probability density function is:
\begin{equation}
    \label{q PDF}
    P(q) = q^{\gamma}, \hspace{5mm} \textrm{for}~q_{\rm min} \leq q \leq 1,
\end{equation}
where $\gamma = -0.4$ and $q_{\rm min} = 0.01$ by default \edit{\citep{2013ApJ...764..155L, KobulnickyFryer2007, KiminkiKobulnicky2012}}.
\edit{Note that many of these parameters and relations are uncertain. We allow all of these parameters which are set by default to be changed by the user.}

The default resolved properties for the \texttt{SPISEA} multiple star system populations are drawn from observations of the local solar neighborhood. 

\cite{2013ARA&A..51..269D} summarize that the frequency of semi-major axes often follows a log normal distribution whose parameters depend on mass, particularly for systems with a lower mass primary star. However, A-type stars were described as having a bimodal distribution and OB-type as a peak with a power law tail. We characterize the separation of multiples as a log-normal distribution with a mean and standard deviation changes with mass. 

We fit the mean semi-major axis vs. mass as a broken power law, 
\begin{equation}
\label{broken_PL}
    a(M) = 
    \begin{cases}
        A(M/M_{\text{break}})^{-\alpha_1} & M < M_{\text{break}} \\
        A(M/M_{\text{break}})^{-\alpha_2} & M > M_{\text{break}} \\
    \end{cases}
\end{equation}
where A = 379.8 AU, $M_{\text{break}}$ = 4.9 $M_{\odot}$, $\alpha_1$ = -1.8, and $\alpha_2$ = 4.2.
We also limit the maximum separation to 2000 AU \citep{2013ARA&A..51..269D}. We fit the standard deviation of the log of the semi-major axis as a line,  
\begin{equation}
\label{sigma_loga}
    \sigma_{log_{10}(a)} = m \; \textrm{log}_{10}(M) + b
\end{equation} 
where $M$ is in $M_{\odot}$, $a$ is in AU, m = 0.84, and b = 0.31
(see Fig.~\ref{fig:a}).
Since the most massive stars are not described by a log normal in \cite{2013ARA&A..51..269D}, we set any objects $M > 2.9 M_{\odot}$ to $\sigma_{\log(a)}(2.9 M_{\odot})$ (where $2.9 M_{\odot}$ is the lower mass limit of A-type stars) since extrapolating upwards in mass leads to excessively large dispersion in $a$.

\edit{For the eccentricity distribution, it is unclear if the binaries follow a Maxwellian ``thermal" distribution (f(e) = 2e) or a flat one (f(e) = constant). A thermal distribution would imply significant interaction between the objects in the binary and their environment so they are able to ``thermalize." A flat distribution would imply more isolated evolution and time to circularize. \cite{2013ARA&A..51..269D} summarize that binaries with $0.5 \textrm{ AU} \lesssim a \lesssim 10$ AU are consistent with a flat distribution. Though it is also possible that there is a mass dependent distribution in which less massive primaries have a flatter eccentricity distribution and more massive primaries have a more thermal one \citep{Moe&DiStefano:2017}. Eccentricities of wider binaries have been measured with Gaia; \cite{Tokovinin:2020} and \cite{Hwang:2022} found closer binaries have a more uniform distribution and wider binaries have a more thermal or even superthermal distribution. In this paper we will use the thermal distribution with $0 \leq e \leq 1$ and will explore more complex eccentricity distributions in future work.}

The rest of the orbital parameters are determined with standard Keplarian distributions. $i$ is the inclination of the system, which is how tilted the system is with respect to the plane. $\Omega$ and $\omega$ describe the orientation of the tilted system in the plane, where $\Omega$ is the angle of the system around the reference plane and $\omega$ is the angle between periapsis and the reference plane in the plane of the system. If $i$ were 90$^{\circ}$, $\omega$ would be perpendicular to the reference plane. The frequency of inclination was assumed to be flat in $cos(i)$, and $\omega$ and $\Omega$ were both assumed to be isotropic. See Table \ref{tab: spisea parameters} for a summary.

The stars in \texttt{SPISEA} are then evolved with an Initial-Final Mass Relation (IFMR) which determines whether a star becomes a white dwarf, neutron star, or black hole. \cite{Rose:2022} explores the impact of metallicity-dependent and independent IFMRs in \texttt{SPISEA} and \texttt{PopSyCLE}. In this paper we use a metallicity-dependent IFMR based on the simulations of \citet{Sukhbold:2014} and \citet{Sukhbold:2016} (called SukhboldN20). We leave the inclusion of binary mass exchange, mergers, or ejections to future work (see Section \ref{sec: future improvements}); however, since stars massive enough to become black holes are all in multiple systems and the objects are evolved independently, the black holes all remain in multiple systems.

\begin{deluxetable*}{l|l|l|l|l}
    \centering
    \tablecaption{Binary table parameters \label{tab: spisea parameters}}
    \tablehead{\colhead{Column} &  \colhead{Description}  & \colhead{Units} & \colhead{\texttt{SPISEA}} & \colhead{\texttt{PopSyCLE}}}
       \startdata
        log(a) & Log of the system semi-major axis & AU & $\times$ & $\times$  \\
        e & 	Eccentricity & --  & $\times$ & $\times$\\
        i & 	Inclination & Degrees  & $\times$ & $\times$\\
        $\Omega$ & 	Longitude of ascending node & Degrees  & $\times$ & $\times$\\
        $\omega$ & 	Argument of periapsis & Degrees  & $\times$ & $\times$\\
        P & Period & Days  &  & $\times$\\
        $\alpha$ & Angle between North and binary axis (East of North) & Degrees &  & $\times$\\
        $\phi_{\pi_E}$ & Angle between North and relative proper motion (East of North) & Degrees  &  & $\times$\\
        $\phi$ & Angle between binary axis and relative proper motion & Degrees  &  & $\times$\\
        \enddata
    \tablecomments{These are the additional parameters added to \texttt{SPISEA} companion and \texttt{PopSyCLE} companion output tables. See Table 2 in \cite{2020AJ....160..143H} for the rest of the columns. $\times$ in the \texttt{SPISEA} column indicates that the parameter is available in \texttt{SPISEA} (see Section \ref{sec: binary params} for more details) and $\times$ in the \texttt{PopSyCLE} column indicates that the parameter is available in \texttt{PopSyCLE} (see Section \ref{subsubsec: bin parameters}  for more details).}
    
\end{deluxetable*}

\subsection{Adding multiple systems to \texttt{PopSyCLE}}
\label{sec: popsycle}
Population Synthesis for Compact-object Lensing Events (\texttt{PopSyCLE}) performs a population synthesis with a combination of stars from \texttt{Galaxia} and compact objects and multiple system companions from \texttt{SPISEA}. \texttt{Galaxia}, \citep[][]{2011ApJ...730....3S} is an implementation of the analytical Besan\c con model for the Milky Way \citep{2003A&A...409..523R}, which simulates single stars. 

\subsubsection{Microlensing Terminology}
\label{terminology}
The Einstein radius is the characteristic angular scale in the microlensing event and depends on the mass of the lens ($M$) and distance between the objects ($d_l$ is the distance between the observer and the lens and $d_s$ is the distance between the observer and the source):
\begin{equation}
    \label{thetaE}
    \theta_E = \sqrt{\frac{4GM}{c^2} \left(\frac{1}{d_l} - \frac{1}{d_s}\right)}.
\end{equation}
The Einstein crossing time is how long it takes for the source to cross the Einstein radius:
\begin{equation}
    \label{tE}
    t_E = \frac{\theta_E}{\mu_{\rm rel}},
\end{equation}
where $\mu_{\rm rel}$ is the relative proper motion between the source and lens.

The Einstein radius and Einstein crossing time are both proportional to the square root of the mass of the lens, so more massive lenses result in longer microlensing events and a larger cross-sectional area. For events with multiple system lenses, $t_E$ can be approximated by using the system mass (primary mass + companion masses) of the lens. For events with multiple system sources, the $t_E$ definition is unchanged (see Table \ref{tab: model definitions} for a detailed breakdown).
The projected heliocentric source-lens separation defined in units of $\theta_E$ is:
\begin{equation}
    u(t) = \sqrt{u_0^2 + \left(\frac{t - t_0}{t_E}\right)^2},
\end{equation}
where $u_0$ is the closest lens-source separation in units of $\theta_E$, $t_0$ is the time of closest approach between the source and lens, and $t$ is the time of observation.
The amplification of the source is a function of $u$:
\begin{equation}
    A(t) = \frac{u^2 + 2}{u\sqrt{u^2 +4}},
\end{equation}
and is maximized when $u = u_0$. The light from the source is often blended with nearby objects in the seeing disk which can contaminate light from the source. The blend source flux fraction is defined as:
\begin{equation}
    b_{sff} = \frac{F_S}{F_S + F_L + F_N}
\end{equation}
where $F_S$ is the flux of the source, $F_L$ is the flux of the lens, and $F_N$ is the flux of the neighbors or nearby objects in the seeing disk. When $b_{sff}$ = 1, the event is totally unblended and flux is only observed from the source, and maximum blending is achieved when $b_{sff}$ approaches zero.

The bump magnitude ($\Delta m$) is the difference between the peak magnitude and baseline magnitude:
\begin{equation}
\label{eq:delta m}
    \Delta m = -2.5log_{10}\left(\frac{A(t_0)F_S + F_L + F_N}{F_S + F_L + F_N}\right).
\end{equation}

These equations do not take into account the motion of the Earth around the Sun. As Earth orbits, our perspective on the event changes. We can measure this microlensing parallax signal defined as: 
\begin{equation}
    \label{piE}
    \pi_E = \frac{\pi_{\rm rel}}{\theta_E}
\end{equation}
where $\pi_{\rm rel}$ is the relative parallax between the source and lens:
\begin{equation}
    \label{pirel}
    \pi_{\rm rel} = (1 \textrm{AU})\left(\frac{1}{d_l} - \frac{1}{d_s}\right).
\end{equation}
This signal may manifest in asymmetry and additional peaks in the light curve.

Combining Eqs.~\ref{thetaE}, \ref{piE}, and \ref{pirel} yields an estimate of the lens mass:
\begin{equation}
    \label{lens-mass}
    M = \frac{\theta_E}{\kappa \pi_E}
\end{equation}
where $\kappa \equiv \frac{4G}{(1~AU)c^2}$. $\pi_E$ is routinely measured with photometric light curves for microlensing events with $t_E>100$ days; however, $\theta_E$ measurements are rare and require astrometric measurements or detection of finite source effects.

In microlensing events, the multiple-system can be the lens, source, or both. These are referred to as point source-binary lens (PSBL), binary source-point lens (BSPL), and binary source-binary lens (BSBL), respectively. When there is no binary,
this is referred to as point source-point lens (PSPL). See Table \ref{tab: model definitions} for additional details.

\subsubsection{Adding Companions to \texttt{PopSyCLE}}
\label{sec: adding multiple stars}
Multiples are added to the population synthesis aspect of the code in two different ways since the stars are from \texttt{Galaxia} and compact objects and injected from \texttt{SPISEA}. For each \texttt{PopSyCLE} age and metallicity bin, a \texttt{SPISEA} cluster is generated with the appropriate mass containing both compact objects and stars with a maximum of two companions. \edit{The IMF of the primaries is from \texttt{Galaxia} which has different IMFs for each of the major stellar populations (e.g. bulge, disk halo) \citep[see][Table 1]{2011ApJ...730....3S}. The matching clusters in \texttt{SPISEA} are drawn from a Kroupa mass function, which is used to populate all compact objects; however, the intrinsic \texttt{Galaxia} mass function for stars is preserved through mass-matching, as described in \cite{2020ApJ...889...31L}, Section 3.} 

As before, the \texttt{SPISEA} compact objects are added to the \texttt{PopSyCLE} population of stars already generated from \texttt{Galaxia}. Any companions to compact objects are also injected from \texttt{SPISEA} into the \texttt{PopSyCLE} population. 

\begin{figure*}
    \includegraphics[scale = 0.25]{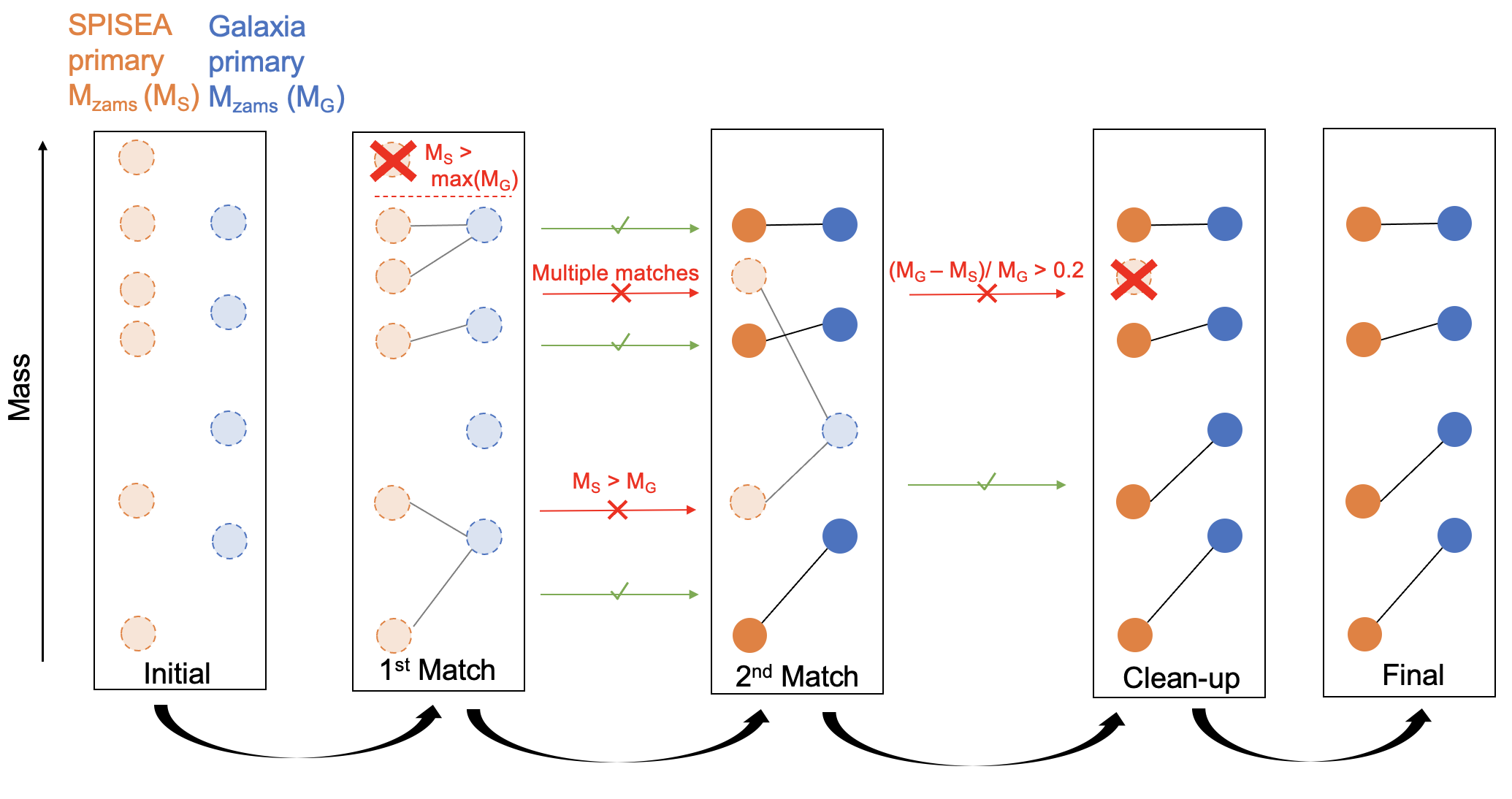}
    \caption{We show the matching process of the \texttt{Galaxia} stars and \texttt{SPISEA} primary stars. The orange circles represent \texttt{SPISEA} primary stars and the blue circles represent \texttt{Galaxia} stars. Circles with a dotted boarder and faint color represent unmatched objects and filled in circles represent objects with a confirmed match. Lines drawn between dotted circles are tentative matches which are either confirmed or rejected indicated with a green line with a checkmark or a red line with an x, respectively. The text above the red line indicates why the match was rejected. The circles are ordered so those with the smallest ZAMS mass are at the bottom and the largest ZAMS mass are at the top. From left to right, the first panel shows the initial configuration of the objects. The second panel shows the first set of matches where the \texttt{SPISEA} stars are matched to the nearest \texttt{Galaxia} star in mass. Any \texttt{SPISEA} stars more massive than the most massive \texttt{Galaixa} star are eliminated. Three of the matches shown are accepted and two are rejected. The upper match is rejected since there are multiple \texttt{SPISEA} stars matched to the same \texttt{Galaxia} star, so the further obect is rejected. The lower match is rejected since the \texttt{SPISEA} mass is larger than the \texttt{Galaxia} mass - this criteria is to ensure that $q$ remains $< 1$. In the third panel, the \texttt{SPISEA} stars that are still unmatched are matched to the next closest \texttt{Galaxia} star in mass. One match is accepted and the other is rejected since the closest \texttt{Galaxia} star mass is further than than 20\% away from the remaining \texttt{SPISEA} star. In the fourth panel, there are no more unmatched \texttt{Galaxia} stars and the final \texttt{SPISEA} star is eliminated. The final configuration of matches is shown on the right.} 
    \label{fig: spisea galaxia match}
\end{figure*}

We then match \texttt{SPISEA} primary stars with \texttt{Galaxia} stars by mass. A summary of this matching process can be seen in Fig~\ref{fig: spisea galaxia match}. 
We do so by making a KDtree of the \texttt{Galaxia} star masses and search the tree iteratively for \texttt{SPISEA} primary mass matches requiring the following:
\begin{itemize}
    \item $M_{\rm Galaxia} > M_{\rm SPISEA}$ to ensure that $q$ remains $<1$
    \item $(M_{\rm Galaxia} - M_{\rm SPISEA})/M_{\rm Galaixa} < 0.2$
    \item Unique matches
\end{itemize}
where $M_{\rm Galaxia}$ is the ZAMS mass of the star from \texttt{Galaxia} and $M_{\rm SPISEA}$ is the ZAMS mass of the primary star from \texttt{SPISEA}. We continue to iteratively search for the nearest neighbors in mass if any of the above criteria are unmet. We remove all systems with $(M_{\rm Galaxia} - M_{\rm SPISEA})/M_{\rm Galaixa} > 0.2$ to the nearest neighbor and systems where $M_{\rm SPISEA} > max(M_{\rm Galaxia})$, which affects 0-5\% of systems.  We then associate the \texttt{SPISEA} companions with the \texttt{Galaxia} stars so that the \texttt{Galaxia} stars replace the \texttt{SPISEA} primary stars. 

The microlensing event calculation aspect of the program is modified in the following ways:
\begin{itemize}
    \item Event $t_E$, $u_0$, $\theta_E$, maximum magnification, and bump magnitude ($\Delta m$) are now calculated using the total system mass defined as $M_{\rm sys} = \sum_{i =1}^{N_{\rm obj}} M_i$ where $N_{\rm obj}$ is the number of objects in the system (i.e. $N_{\rm obj}$ = 3 for triples). 
    \item  All magnitudes in primary and event tables are system luminosities that include flux contributions from the primary and any companions for the system. The system luminosity is used to calculate the $b_{sff}$ values.
    \item The search radius is increased from n$\theta_E$ (where n is a user specified parameter defaulting to 2) to (n$\theta_E$) + $a\sin(i)$ (where $a\sin(i)$ is the projected binary separation). \texttt{PopSyCLE} continues to use the position of the primary for the center of the search radius for candidate events.  
    Since the projected separation between the binary lenses was found to be larger than $\theta_E$ approximately 50\% of the time, we extend the candidate radius to find all potential events.
\end{itemize}

\subsubsection{Available Binary Parameters}
\label{subsubsec: bin parameters}

We calculate binary parameters for each companion in every system including multiple lenses and sources. The following parameters are calculated so binary light curves can be simulated later: the current mass ratio between the companion and primary star ($q$), projected separation in milliarcseconds, period in years, and three binary angles: $\alpha$, $\phi_{\pi_E}$ and $\phi$ (see explanation below and Eqs.~\ref{eq: alpha}-\ref{eq: phi}). 

The period in years is calculated using Kepler's third law:
\begin{equation}
\label{eq: period}
    P = \sqrt{\frac{4\pi^2a^3}{GM}},
\end{equation}
where $a$ is the separation in AU and $M$ is the total mass of the system in $M_{\odot}$.

$\alpha$ is the angle between the binary axis and North (East of North). It is calculated as:
\begin{equation}
\label{eq: alpha}
    \alpha 
    = \textrm{tan}^{-1}\left(\frac{\Delta z_{\rm RAcos(Dec)}}{\Delta z_{\rm Dec}}\right)  - 180^{\circ}
\end{equation}
where $\vec{\Delta z}$ = ($\Delta z_{\rm RAcos(Dec)}$, $\Delta z_{\rm Dec}$) 
is the position of the companion with respect to the primary
as calculated in Section \ref{subsubsec: orbital mechanics} and $\tan^{-1}$ is the \texttt{arctan2} function which returns angles between -$\pi$ and $\pi$ radians.
180$^{\circ}$ is subtracted off to point the vector towards the primary instead of towards the companion.

$\phi_{\pi_E}$ is the angle between the vector of North and the relative proper motion between the source and the lens ($\vec{\mu}_{rel}$), measured as East of North. This is calculated as:
\begin{equation}
\label{eq: phi_piE}
    \phi_{\pi_E} 
    = \textrm{tan}^{-1}\left(\frac{\mu_{\rm rel,RAcos(Dec)}}{\mu_{\rm rel,Dec}}\right)
\end{equation}

$\phi$ is the angle between the binary axis and the proper motion vector where the binary axis points towards the primary. It can be calculated from $\alpha$ and $\phi_{\pi_E}$:
\begin{equation}
\label{eq: phi}
    \phi = \alpha - \phi_{\pi_E}
\end{equation}

\subsubsection{Orbital Mechanics}
\label{subsubsec: orbital mechanics}
We assume a static binary over the entire microlensing survey window (i.e. no orbital motion). The static position is sampled from the orbit using a random orbital phase. We calculate $\vec{\Delta z}$, the position of the companion with respect to the primary using: the orbital parameters $\omega$, $\Omega$, $i$, and $e$ (see the end of Section \ref{sec: binary params}); the orbital period ($P$), the semi-major axis ($a$), and the mass of the primary ($M$); and an initial time $t_{\rm init}$ = nP + $t_0$, where n is a random number between 0 and 1. Following the equations outlined in \cite{1967pras.book.....V} and applied in \cite{2009ApJ...690.1463L}, we first calculate the following Thiele-Innes constants:
\begin{eqnarray}
    A &=& a(\cos \omega \cos \Omega - \sin \omega \sin \Omega \cos i) \\
    B &=& a(\cos \omega \sin \Omega + \sin \omega \cos \Omega \cos i) \\
    F &=& a(-\sin \omega \cos \Omega - \cos \omega \sin \Omega \cos i) \\
    G &=& a(-\sin \omega \sin \Omega + \cos \omega \cos \Omega \cos i)
\end{eqnarray}
The eccentricity anomaly ($E$), which describes where the object is in an elliptical orbit at $t_{\rm init}$, is calculated numerically \citep[see][]{2009ApJ...690.1463L}. Then it is used to calculate the elliptical rectangular coordinates:
\begin{eqnarray}
    X &=& \cos E - e \\
    Y &=& \sqrt{1 - e^2} \sin E.
\end{eqnarray}
The derivative of E is:
\begin{equation}
    \dot{E} = \frac{\bar{m}}{1 - e \cos E},
\end{equation}
where $\bar{m}$ is the mean motion in radians per year:
\begin{equation}
    \bar{m} = 2\pi/P.
\end{equation}
These are then used to find the projected positions and velocities with respect to the primary:
\begin{eqnarray}
    z_{\rm E} &=& BX + GY \\
    z_{\rm N} &=& AX + FY
\end{eqnarray}
where $z$ is in AU.
When modeling lightcurves we assume a static binary. In the future we will incorporate orbital motion into the microlensing event; however, this would significantly add to run time (see Section \ref{sec: future improvements}).

\subsubsection{Binary Light Curves}
\label{sec: bin lightcurves}
Previously in \texttt{PopSyCLE}, the user would determine whether an event is detectable using the baseline magnitude, $u_0$, and $\Delta m$ at closest approach. However, for binary and triple systems, the point of maximum magnification may not correspond to the closest approach to the primary. To properly determine detectability for multiple systems, we simulate lightcurves for all the events that contain a multiple system and store the parameters. In the case of triple lenses/sources, we simulate multiple binary lightcurves with each primary-secondary pair and choose the one with the largest amplitude (see Appendix \ref{sec: model definitions} for details). 
See Appendix \ref{appendix: triples} for an analysis on the relative effect of triple systems, but in this analysis, we leave fully simulating the triple systems to future work (see Sec.~\ref{sec: future improvements}).

The following parameters were then calculated for each event:
\begin{enumerate}
    \item Number of peaks ($n_p$) in the light curve defined with a prominence of at least $10^{-5}$ magnitudes by scipy \texttt{find\_peaks()}. The prominence is defined as the minimum height needed to descend from the peak until a higher peak's baseline is reached, where a peak is a local maxima in magnitude space (lowest magnitude).  $10^{-5}$ is far in excess of what is detectable by current ground based observatories; however, we cut based on this later and chose this value to allow for future flexibility.
    \item Bump magnitude defined as the maximum magnitude - minimum magnitude. For an event with a single peak, the prominence is equivalent to the bump magnitude. 
    \item Time at which the highest peak occurs. 
    \item Average of the $n_p$ times at which the peaks occur. 
    \item Standard deviation of the $n_p$ times at which the peaks occur. 
\end{enumerate}
For events with more than one peak, we calculate the following parameters for each peak in a separate table:
\begin{enumerate}
    \item Time at which the peak occurs. 
    \item Bump magnitude of that peak, defined as the maximum magnitude of the peak - minimum magnitude of peak. ($\Delta m$)
\end{enumerate}
For additional parameters see \texttt{PopSyCLE} documentation.\footnote{\url{https://popsycle.readthedocs.io/en/latest/}}

\section{ \texttt{PopSyCLE} Simulations}
\label{sec: computation}

We ran \texttt{PopSyCLE} simulations for eighteen fields each with area 0.34 deg$^2$. As we compare the results to the OGLE IV Microlensing Survey \citep{ogleIV:Udalski:2015}, we chose fields to correspond with OGLE IV fields which are listed in Table \ref{tab:fields run}. \edit{The simulations are publicly available.\footnote{\url{https://w.astro.berkeley.edu/popsycle/}}}

These fields were simulated 
in the I filter with a 0.65" blend radius to mimic OGLE's median seeing of 1.3" \citep[see Section 4 of][]{2020ApJ...889...31L}. The Galactic parameters used correspond to \texttt{Galaxia} v3 \citep[see Appendix A of][]{2020ApJ...889...31L}. We did two sets of simulations where we will call runs with multiples M Runs (multiples runs) and runs without S Runs (singles runs).

\begin{deluxetable}{c|c|c}
    \centering
    \tablecaption{Fields run in \texttt{PopSyCLE} \label{tab:fields run}}
    \tablehead{\colhead{Field Name} &  \colhead{$\ell$}  & \colhead{$b$}}

       \startdata
        BLG 500 & 	1.00$^{\circ}$ & -1.03$^{\circ}$ \\
        BLG 501 & 	-0.61$^{\circ}$ & -1.64$^{\circ}$ \\
        BLG 504 & 	2.14$^{\circ}$ & -1.77$^{\circ}$ \\ 
        BLG 505 & 	1.09$^{\circ}$ & -2.39$^{\circ}$ \\
        BLG 506 & 	-3.00$^{\circ}$ & 0.01$^{\circ}$ \\
        BLG 507 & 	-1.06$^{\circ}$ & -3.61$^{\circ}$ \\
        BLG 511 & 	3.28$^{\circ}$ & -2.52$^{\circ}$ \\
        BLG 512 & 	2.22$^{\circ}$ & -3.14$^{\circ}$ \\
        BLG 527 & 	8.81$^{\circ}$ & -3.64$^{\circ}$  \\
        BLG 534 & 	-1.14$^{\circ}$ & -2.25$^{\circ}$ \\
        BLG 535 & -2.21$^{\circ}$ & -2.86$^{\circ}$ \\
        BLG 629 & 	7.81$^{\circ}$ & 4.81$^{\circ}$ \\
        BLG 611 & 	0.33$^{\circ}$ & 2.82$^{\circ}$ \\
        BLG 645 & 3.20$^{\circ}$ & -1.15$^{\circ}$ \\
        BLG 646 & 	4.34$^{\circ}$ & -1.90$^{\circ}$ \\
        BLG 648 & 	1.96$^{\circ}$ & 0.94$^{\circ}$ \\
        BLG 652 & 	1.85$^{\circ}$ & 2.30$^{\circ}$ \\
        BLG 675 & 	0.78$^{\circ}$ & 1.69$^{\circ}$
        \enddata   
    \tablecomments{This is the list of fields simulated and analyzed. The first column is the name, which ``BLG \#" correspond to the name "OGLE-IV-BLG\#". The second column is galactic longitude, $\ell$. The third column is galactic latitude, $b$. Each field has an area of $0.34~\textrm{deg}^2$.}
\end{deluxetable}

\edit{For the simulations, we used extinction as described in \cite{2020ApJ...889...31L} which uses the \cite{Schlegel:1998} E(B-V) values from which \texttt{Galaxia} produces an estimated integrated 3D dust map and the \cite{Damineli:2016} reddening law. See \cite{2020ApJ...889...31L} Appendix B for CMDs compared to OGLE.}

Microlensing events were identified using observability criteria to match the limits of OGLE observations. See Table \ref{tab:cuts} for a summary of cuts that were made to mock results. Unless otherwise specified, we use Mock OGLE Early Warning System \citep[EWS,][]{ogleIV:Udalski:2015} cuts. For OGLE EWS we required $I_{\rm base} \leq 21$, an $I$-band bump magnitude of $\geq 0.1$, and $\edit{|}u_0\edit{|} \leq 2$. 
The bump magnitude ($\Delta m$) cut is made to remove low amplitude events \citep{2020ApJ...889...31L}. For events containing a multiple system, the bump magnitude was calculated from the simulated binary light curves (see Section \ref{sec: bin lightcurves}). Multiple-lens events are only counted as multi-peak if the secondary peak has $\Delta m \geq 0.02$. This is an estimate based on the mean OGLE uncertainty \citep[see Fig.~2 of][]{2016AcA....66....1S}, since if a primary peak is identified, the light curve will be scrutinized more carefully for secondary peaks. \edit{This does not take into account gaps from cadence, though. In order to do so, we fit the microlensing events with an OGLE-like cadence as described in Section \ref{sec:fit lightcurves}.}
\begin{deluxetable}{c|c|c|c}
    \centering
    \tablecaption{Mock Microlensing Cuts \label{tab:cuts}}
    \tablehead{\colhead{} &  \colhead{Mock EWS}  & \colhead{Mock Mr{\'o}z17} & \colhead{Mock Mr{\'o}z19}}
    \startdata
        Filter & I & I & I \\
        Source Mag & -- & 	$\leq 22$ & $\leq 21$ \\
        Baseline Mag & $\leq 21$ & 	-- &  --  \\
        $\Delta m$ & $\geq 0.1$ &	-- & -- \\
        $\edit{|}u_0\edit{|}$ & $\leq 2$ &	$\leq 1$ & $\leq 1$ 
    \enddata    
    \tablecomments{This table is a summary of the cuts made to simulate the OGLE Early Warning System (EWS), Mr{\'o}z 17 \citep{2017Natur.548..183M}, and Mr{\'o}z 19 \citep{2019ApJS..244...29M}.
    When comparing specifically with OGLE numbers, we make the additional cut of one observable peak, where the second peak must have $\Delta m \geq 0.02$ to be considered observable.}
    
\end{deluxetable}

\section{Fitting Simulated Lightcurves}
\label{sec:fit lightcurves}

\edit{As a post-processing step, we generated OGLE-like photometry for the 5,898 microlensing events that passed Mock EWS cuts (see Table \ref{tab:cuts}), including both PSPL and multiple events, and fit them with PSPL models using \texttt{BAGLE} (Bayesian Analysis of Gravitational Lensing Events)\footnote{\url{https://github.com/MovingUniverseLab/BAGLE_Microlensing/tree/main}}.} 

\edit{We created realistic temporal sampling for each \texttt{PopSyCLE} light curve 
by selecting a real OGLE lightcurve from 2016 from each field from the OGLE EWS webpage and giving all the lightcurves from that field the cadence that matched that real lightcurve. There were no lightcurves alerted for field BLG 629 in 2016, so we chose a lightcurve from a neighboring field, BLG 630 for those lightcurves. We shifted the simulated $t_0$ values from \texttt{PopSyCLE} (which are between -500 and 500) by 56400 days, which is the midpoint in MJD of the avaliable timeseries of the lightcurves. }
\edit{We generated errorbars for the lightcurves by fitting the errors of 100 randomly selected lightcurves from the dataset provided in \cite{2019ApJS..244...29M}. We found the following best fit flux-mag relation and flux errors:
\begin{eqnarray}
    F &=& 350~\textrm{counts} \times 10^{\frac{m - 16}{-2.5}} \\
    F_{\rm err} &=& F^{0.85}
\end{eqnarray}
We also added Gaussian scatter with an amplitude the size of the errorbars.}

\edit{Next we fit the simulated lightcurves with point-source, point-lens models. We fit the events to assess, in more detail, what fraction of multiple events may be mistaken for PSPL and to understand biases in parameters that may be caused by fitting events with multiples with PSPL models. We do not include parallax to match the methodology of the current ground based surveys which do not fit parallax except when the lightcurves fit poorly by a PSPL without parallax fit.  We use priors as described in Table \ref{tab:priors}. We used \texttt{BAGLE} to do the fitting which uses \texttt{pymultinest} nested sampling algorithm. This can be beneficial in a degenerate parameter space like microlensing in which one can often have multimodal solutions. Another benefit to using this fitting algorithm is that we have full posteriors for the microlensing events we fit which can be used for future statistical studies.}

\begin{deluxetable}{c|c}
    \centering
    \tablecaption{Simulated Event Priors \label{tab:priors}}
    \tablehead{\colhead{Parameter} &  \colhead{Prior}}
    \startdata
        $t_E$ & $\mathcal{U}(0, 400)$ \\
        $t_0$ & $\mathcal{U}(55200, 57700)$ \\
        $u_0$ & $\mathcal{U}(-1, 1)$\\
        $m_{\rm base}$ & $\mathcal{U}(m_{\rm base, sim} - 0.5, m_{\rm base, sim} + 0.5)$\\
        $b_{sff}$ & $\mathcal{U}(0.001, 1.25)$
    \enddata    
    \tablecomments{\edit{These are the priors used for fitting the lightcurves that passed the OGLE EWS cuts using \texttt{BAGLE}, the procedure of which is described in Section \ref{sec:fit lightcurves}. We chose the $t_0$ ranges to be the approximate minimum and maximum dates in MJD of the real OGLE EWS timeseries we used. We allow $b_{sff}$ to be above 1 to accommodate for ``negative blending" which can occur due to systematics in background subtraction \citep[e.g.][]{Smith:2007}.}}
\end{deluxetable}

\edit{Once we fit the lightcurves, we eliminated those that did not deviate significantly from a ``straight line fit" (the sigma-clipped ($3\sigma$) mean of the lightcurve). This was determined by calculating the reduced $\chi^2_{\rm const}$ values for each lightcurve and eliminating those that were below the 99th percentile of the corresponding reduced $\chi^2$ distribution. In equation form, lightcurves that met the following criteria were eliminated:
\begin{equation}
    \frac{\chi^2}{dof} < \frac{\chi^2_{0.99}}{dof}.
\end{equation}
where $dof$ is the degrees of freedom in the model.}
\edit{This eliminated 72\% of lightcurves. These lightcurves either had an event that peaked in a seasonal gap or were faint and had a larger amplitude of scatter than the change in magnitude due to the microlensing event.}
\edit{We then determined which fit well by a PSPL lightcurve. By fit well, this means those with a reduced $\chi^2_{\rm PSPL}$ less than the 99th percentile of the corresponding reduced $\chi^2$ distribution. We will discuss these results in Section \ref{sec:fitting results}.}

\section{Results}
\label{sec:results}

\subsection{Fraction of Multiple Events}
\label{subsec: fractions}
\begin{figure*}
    \centering
    \includegraphics[scale = 0.33]{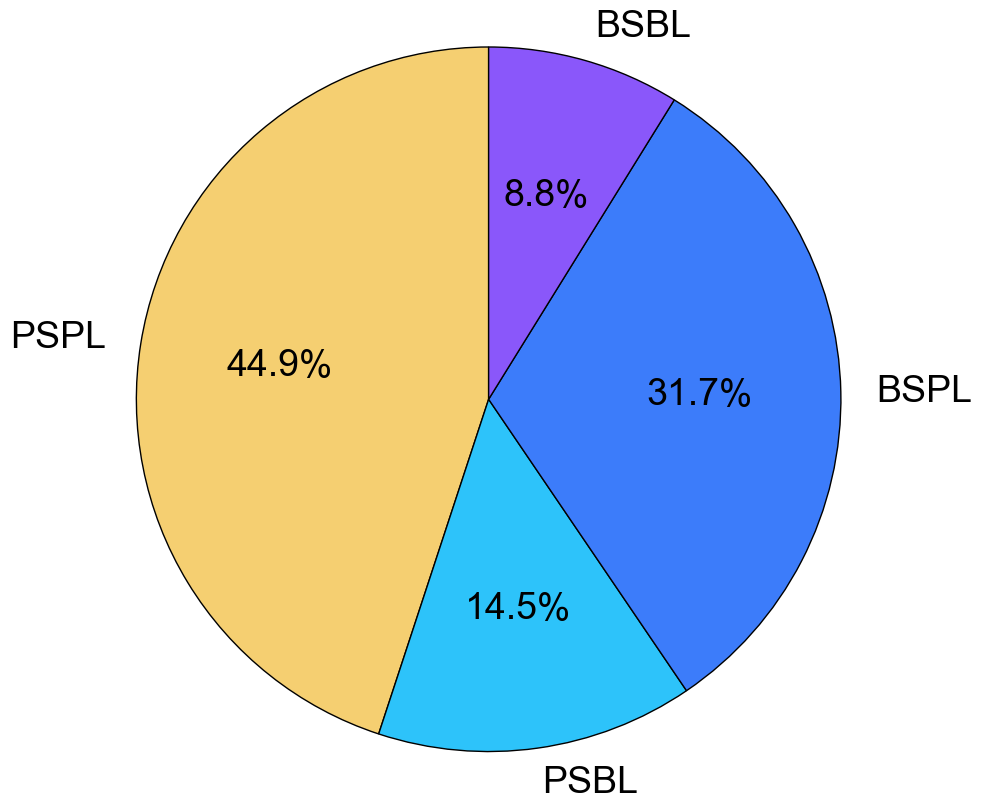}
    \includegraphics[scale = 0.33]{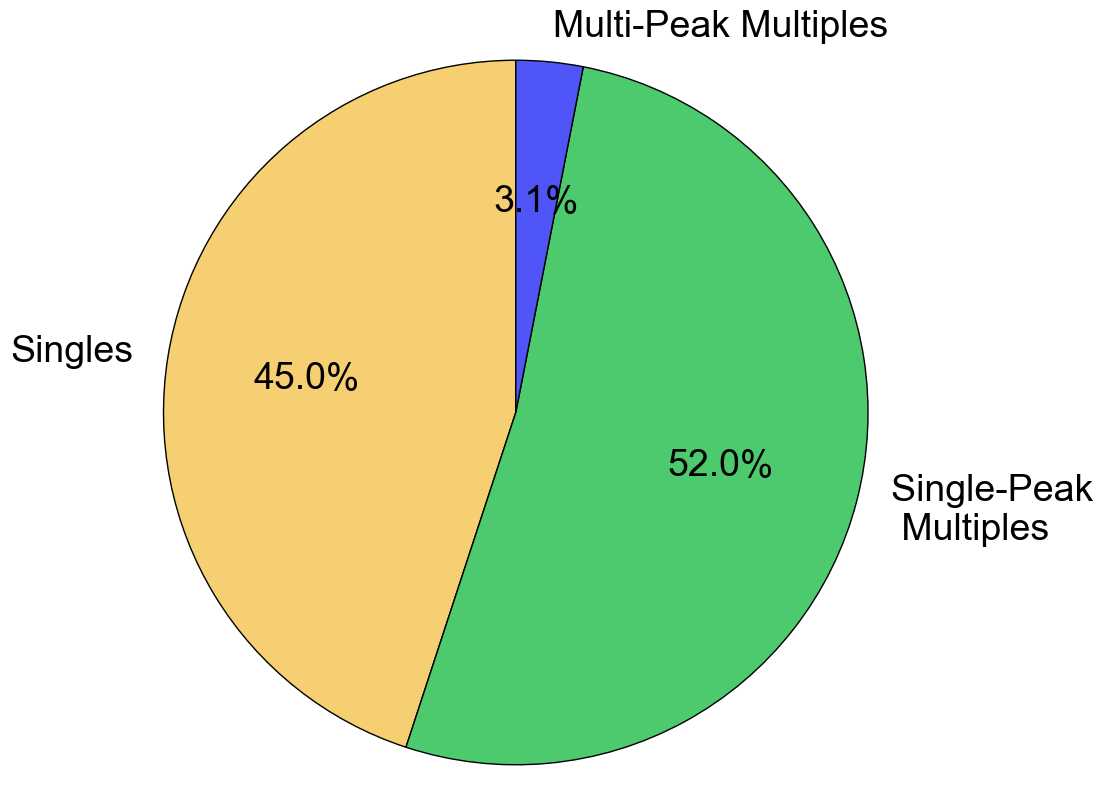}
    \caption{\textit{Left} shows the percentage of the intrinsic population with OGLE EWS observability cuts with a single source and single lens (PSPL), a single source and multiple lens (PSBL; 19\% of which have a triple lens), a multiple source and single lens (BSPL; 15\% of which have a triple source), and multiple source and multiple lens (BSBL; 31\% of which have a triple lens or source) as indicated in Table \ref{tab:binary breakdown}, where PSPL events are in yellow and events with a multiple system are in varying shades of blue. We find that more than half of microlensing events include at least one multiple system. \edit{This fraction can change depending on the minimum magnitude the survey is sensitive to, the wavelength of observation, and the resolution of the survey (see Section \ref{subsec: fractions}).} \textit{Right} shows the percentage of single (PSPL) events, single-peaked events with at least one multiple system (19\% of which have a triple system), and multi-peaked events with at least one multiple system (20\% of which have a triple system). We find that most multiple microlensing events have a single peak and have the possibility of being mistaken as a single-only event.}
    \label{fig:pie chart}
\end{figure*}

We found that over half of our simulated microlensing events contain a multiple lens, source, or both. 
\edit{These are events that passed Mock-EWS cuts (see Table \ref{tab:cuts}) and should be compared with completeness-corrected OGLE EWS results. The results from the sample of fitted events that pass observability criteria before completeness correction will be discussed in Section \ref{sec:fitting results}.} 
As can be seen in the left plot of Figure \ref{fig:pie chart} and Table \ref{tab:binary breakdown}, 55\% of events \edit{that meet Mock OGLE EWS cuts} contain a multiple lens or source. Of those, 94.4\% of multiple events are single peaked and have the potential of masquerading as a PSPL microlensing event. This leaves 3.1\% of all events as obvious, multi-peaked multiple events, as can be seen in the right plot of Fig.~\ref{fig:pie chart}. 10.4\% of all events and 10.1\% of single peaked events have at least one triple system; see the caption of Fig.~\ref{fig:pie chart} and Appendix \ref{appendix: triples} for details. 

\edit{We also investigated the flux ratios of the binaries to probe the fraction with a secondary component that contributes significantly to the flux. We found that 37\% have a fainter component which contributes at least 10\% of the system flux and 6\% have a fainter component which contributes at least 40\% of the flux. Hence, most systems have one component of the binary source which is significantly brighter than the other and is consistent with Fig.~\ref{fig:pie chart} and Table \ref{tab:binary breakdown}, indicating that most binary source events appear as singles. While this is true for an OGLE style sample, for upcoming telescopes such as Roman or Rubin which probe to greater depths, these fractions will likely change.
}

\edit{In addition, we find that the fraction of multiple events is dependent on the minimum magnitude the survey is sensitive to. As the magnitude increases and we include fainter stars, the multiple event fraction decreases by a few percent. The multiple event fraction may also be dependent on the wavelength of observation and the resolution, cadence, and photometric precision of the survey. These effects will be studied in future work.}

\begin{figure*}
    \hspace*{-1.5cm}
    \centering
    \includegraphics[scale=0.64]{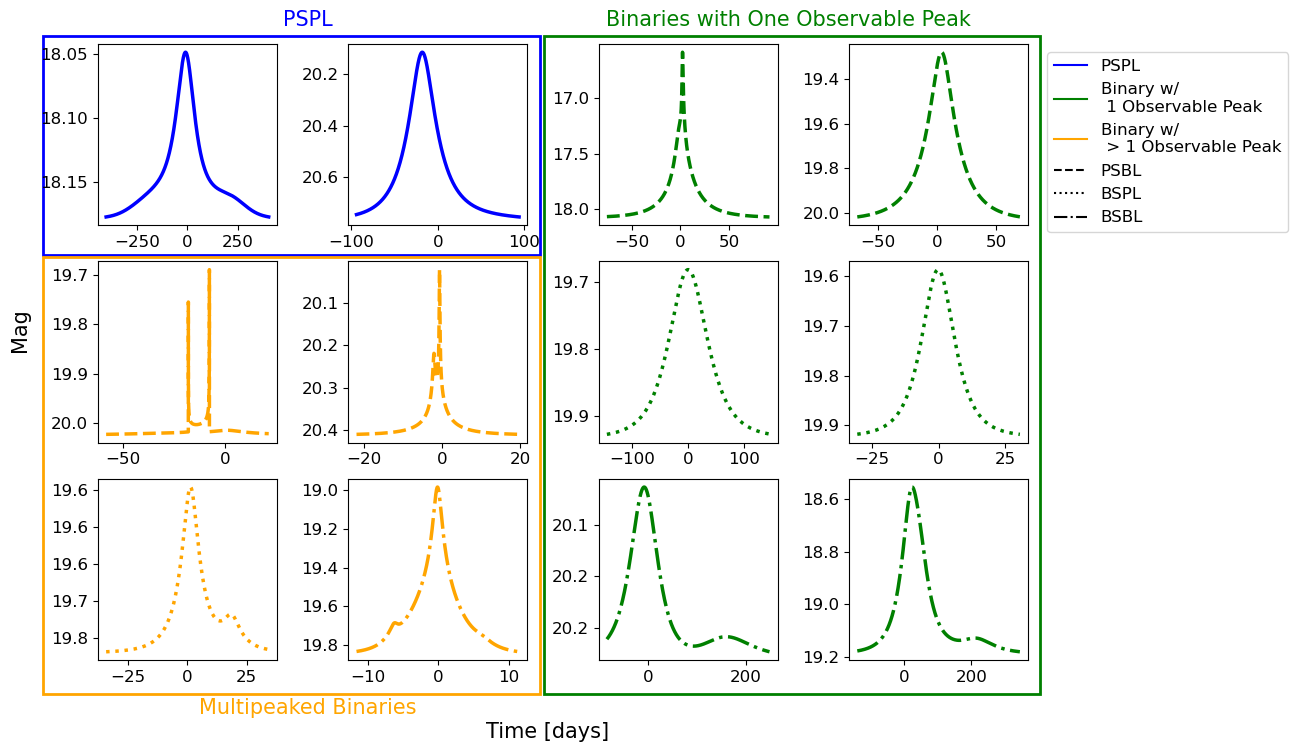}
    \caption{This is a set of random example lightcurves from the \texttt{PopSyCLE} simulation where time is in days where 0 corresponds to $t_0$. In blue are PSPL events. Deviations from the prototypical microlensing shape in these lightcurves are due to parallax. Binary events with more than one peak are plotted in orange and binary events with one observable peak are plotted in green. Dashed lines indicate PSBL, dotted lines indicate BSPL, and dash-dot lines indicate BSBL. As can be seen in the green plots, some binary events have shoulders or small asymmetries, but others appear indistinguishable from typical PSPL events and therefore may masquerade as PSPL events.}
    \label{fig:sample microlensing events}
\end{figure*}

In Fig.~\ref{fig:sample microlensing events}, we plot randomly selected example microlensing lightcurves classified as true singles, single-peaked multiples, and multi-peaked multiples. The green lightcurves in Fig.~\ref{fig:sample microlensing events} are events with a binary source or lens with a single observable peak. While some have a shoulder or some asymmetry, the second peak does not rise above $\Delta m \geq 0.02$ and others appear visibly indistinguishable from typical PSPL events. There may also be parallax signal in PSPL lightcurves (see the left blue curve in Fig.~\ref{fig:sample microlensing events}) which may not fit well by a PSPL-no parallax model. Without modeling work that takes into account these higher order effects, one cannot interpret these populations accurately. 

\begin{deluxetable}{l|l|l|l|l}
    \centering
    \tablecaption{Breakdown of Multiple Microlensing Events \label{tab:binary breakdown}}
    \tablehead{
       \colhead{ }\vspace{-0.3cm} &  \colhead{Both}  & \colhead{Lens} & \colhead{Source} & \colhead{Neither} \\
       \colhead{ } & \colhead{Multiple} &
       \colhead{Multiple} &
       \colhead{Multiple} &
       \colhead{Multiple}}
    \startdata
\% of Total &8.8\%&14.5\%&31.7\%&44.9\% \\ 
Multipeak \% of Total &0.3\%&2.6\%&0.2\%&- \\ 
Multipeak \% of Category &3.5\%&17.8\%&0.5\%&- 
    \enddata
    \tablecomments{This table has the percent of all events where either both the lens and source are multiple systems, just the lens is a multiple system, just the source is a multiple system, or neither are with mock OGLE EWS cuts. `\% of Total' is the percentage of total which are multiple systems. `Multipeak \% of Total' is the percentage of all events which have multiple peaks from that category; i.e. 2.6\% of all events are multi-peaked PSBL events. `Multipeak \% of Category' is the percentages of each category which have multiple peaks; i.e. 17.8\% of PSBL events are multi-peaked. We find the percentage of single-only events with multiple peaks (which are caused by parallax) to be negligible ($\sim 0.04\%$).
    }
\end{deluxetable}

\begin{figure}
    \centering
    \includegraphics[width=0.48\textwidth]{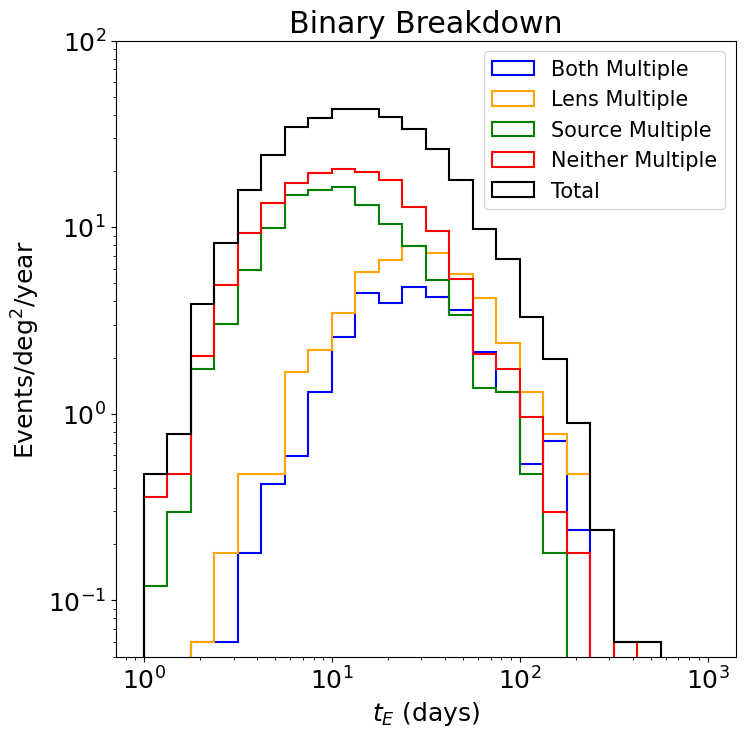}
    \caption{Number of events/deg$^2$/year as a function of Einstein crossing time (Eq. \ref{tE}) in days for M Runs. The black distribution is the total, the blue is events where both the lens and source are multiples, yellow is events where only the lens is a multiple, source is events where only the source is a multiple, and red is events where neither are multiples. In general, events with a multiple lens have longer Einstein crossing times than those which do not, which makes sense since companions add mass leading to longer $t_E$ if the mass is added to the lens. See Table \ref{tab:binary breakdown} for the percent of each population and Table \ref{tab:properties} for the mean and median $t_E$.}
    \label{fig:binary breakdown}
\end{figure}

We can also see in Table \ref{tab:binary breakdown} that most multipeaked events are PSBL (2.6\% of all events). Also, a higher percentage of PSBL events are multi-peaked. 17.8\% of PSBL events are multi-peaked, 0.5\% of BSPL events are multi-peaked, and 3.5\% of BSBL events are multi-peaked. Our simulated percentages of multi-peaked events matches well with the fraction of binary events detected by surveys, with 2-11\% of events being determined to be binaries. Most early surveys identified binaries by visual inspection and found 6\% binaries \citep{Alcock:2000-Binaries}, 3-11\% binaries \citep{Jaroszynski:2002}, 1.8\% binary lenses and 1.8\% binary sources \citep{Jaroszynski:2004}, 3.1\% binary lenses and 0.8\% binary sources \citep{Jaroszynski:2006}, 1.5\% binary lenses and 0.7\% binary sources \citep{Skowron:2007}, 1.5\% binary lens and 0.1\% binary sources \citep{Jaroszynski:2010}. More recent studies have cut out ``anomalous" events (i.e. events with large binary and parallax signals) which do not fit well by a PSPL-model with no parallax with a $\chi^2$ criteria (i.e. $\chi^2/dof \leq 2$ in \citealp{2019ApJS..244...29M}). They then scale up the event rates and optical depths by a factor to account for those which have been removed based on previous established fractions of binary events. They scale assuming a binary fraction of 6\% and that contribution to the optical depth of a binary lens is $\sqrt{2}$ more than a single lens \citep{Sumi:2006, Sumi:2013, 2019ApJS..244...29M}, though \cite{2019ApJS..244...29M} mention they find $\sim 10\%$ anomalous events. While we can reproduce the fraction of ``obvious" binaries that are caught by surveys, this still indicates there is a significant fraction of events with multiples in data which are masquerading as PSPL events. Events with multiple lenses tend to have longer $t_E$ values than those without (see Fig.~\ref{fig:binary breakdown} and Table \ref{tab:binary breakdown}), so when interpreting $t_E$ distributions from photometric surveys, it is critical to take multiples into account.

\subsection{Comparison of \texttt{PopSyCLE} to OGLE data}
\label{sec: ogle data}
Adding multiples drove the mean of the Einstein crossing time distribution to a longer time and also led to an increased population of events at long $t_E$. The mean $t_E$ increased by 2.6 days and the median $t_E$ by 1.2 days
(see Table \ref{tab:properties}). 
A number of these binaries, though, would be unlikely to be mistaken for PSPL events. Those with multiple observable peaks 
can be cut out.  When comparing the single peaked population to singles only, the mean $t_E$ increased by 2.2 days and the median $t_E$ increased by 0.9 days.

\begin{figure}
    \centering
    \includegraphics[width=0.48\textwidth]{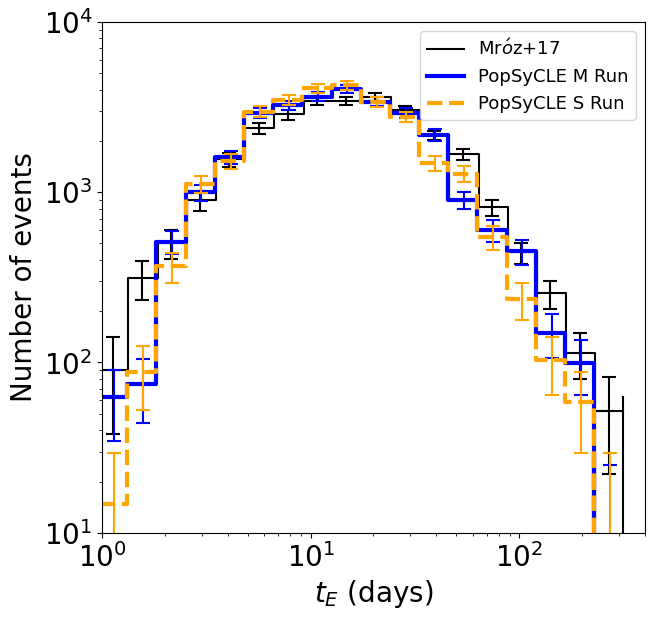}

    \caption{Number of events as a function of the Einstein crossing time in days with the distribution of events from S Runs in the yellow dashed line, single and single-peaked multiple events from M Runs in the blue solid line, and event distributions from \cite{2017Natur.548..183M} divided by their appropriate detection efficiencies. The simulations are both with OGLE-M17 style cuts (see Table \ref{tab:cuts}). The S Runs and M Runs are each scaled such that the number of events are the same as the data. Only single and single-peaked multiple events are kept for the M Runs since OGLE only includes those they fit as PSPL events in their analysis. The \texttt{PopSyCLE} data is from a subset of fields simulated in Table \ref{tab:fields run} to match those in \cite{2017Natur.548..183M} -- they report fields 500, 501, 504, 505, 506, 511, 512, 534, and 611. Neither distribution matches well at the low-$t_E$ end, which is expected due to the lack of brown dwarfs in \texttt{PopSyCLE}. The M Runs match the peak and the long duration events of the OGLE distribution within the errorbars in most cases, more than in the S Run case.}
    \label{fig:mroz17}
\end{figure}

To compare our $t_E$ distribution with OGLE, we used \edit{the} distributions in \cite{2017Natur.548..183M} from Extended Data Table 4 corrected with detection efficiencies from Extended Data Table 5. We took the 9 \edit{simulated \texttt{PopSyCLE}} fields that correspond to the \edit{OGLE} fields analyzed in that paper and made Mock Mr{\'o}z17 cuts to match. With single systems only, the peak of \texttt{PopSyCLE}'s $t_E$ distribution was less than that of the data and \edit{the distribution was} unmatched at long $t_E$ values. The presence of multiple systems in the simulation brought our simulated $t_E$ distribution into better agreement with the observed distribution. We can see this shifted the \texttt{PopSyCLE} distribution into better alignment with the peak and at long durations with the OGLE distribution (see Fig.~\ref{fig:mroz17}). Since the \texttt{PopSyCLE} simulation does not include brown dwarfs or exoplanets, we do not expect the distributions to match at $t_E \lesssim 3$ days.

We performed a Kolmogorov-Smirnov two-sample test between the data from \cite{2017Natur.548..183M} and the M Run and S Run data in Fig.~\ref{fig:mroz17}. We made 1000 realizations of the OGLE distribution and found a mean p value of $9.34 \times 10^{-4}$ when comparing M Runs and OGLE and $3.06 \times 10^{-10}$ when comparing S Runs and OGLE.

Adding in multiples also brought the observed field-averaged event rate into better agreement with our simulations. The average between fields can be seen in Table \ref{tab:mroz19_comparison_field_avg} and the individual fields in Table \ref{tab:mroz19_comparison}. We find that the event rate in events/deg$^2$/year is closer and within the errorbars of the data. The $N_{\rm stars}/\textrm{deg}^2$ data has better agreement with M Runs, but events/star/year has better agreement between S Runs and the data (though they are similar within the errorbars), assuming each multiple star system is blended as one star; however, there is confusion in the number of stars. If stars are blended, this will artificially increase their brightness, which could bring them above the magnitude cut, but also blending will decrease the number of stars that can be distinguished. OGLE efficiency corrects their $N_{stars}$ with the Hubble Space Telescope (HST) as truth \citep{Holtzman:2006}, but we do not attempt to find the number of stars as measurable by HST (where multiple star systems are counted a single blended star), so events/deg$^2$/year is the more comparable measurement. \edit{See \cite{2020ApJ...889...31L} Appendix A and B for additional comparisons between \texttt{PopSyCLE} and data.}

\begin{deluxetable}{c|c|c|c}
    \centering
    \tablecaption{\texttt{PopSyCLE} Statistics \label{tab:properties}}
       \tablehead{\colhead{Type} &  \colhead{$\langle t_E\rangle$~(d)}  & \colhead{med($t_E$) (d)} & \colhead{$f_{BH}$~at~$t_E > $~120d}}
       \startdata
S~Run &19.1&12.8&0.31 \\ 
M~Run total &21.7&14.0&0.29 \\ 
M~Run Single Peak &21.3&13.7&0.31 \\ 
M~Run Sing Lens &16.8&11.3&0.00 \\ 
M~Run Mult Lens &37.8&27.8&0.38 

    \enddata
    \tablecomments{This table has the mean $t_E$ in days, median $t_E$ in days, and fraction of lenses that are black holes with $t_E>$ 120 days for a variety of runs and cuts. The first row are the values for S Runs and the rest are for various populations of the M Runs. Each population has Mock OGLE EWS cuts applied as described in Table \ref{tab:cuts}.}
\end{deluxetable}

\begin{deluxetable*}{c|c|c|c|c|c|c}
    \centering
    \tablecaption{This table compares the field average of singles-only simulations (S Runs) and single-peaked multiples+singles simulations (M Runs) to \cite{2019ApJS..244...29M} values by making the cuts indicated in Table \ref{tab:cuts}. We average the results across the fields we simulated for the \cite{2019ApJS..244...29M} results. We compare the number of stars/deg$^2$ (where multiple star systems are counted a single blended star), events/star/year, events/deg$^2$/year and the mean $t_E$ between 0.5 and and 300 days. We also compute the median $t_E$ between 0.5 and 300 days. $N_{\rm stars}$ is $10^{-6}$ the value in the simulation. We take the average of each of the fields where the errors are the error on the mean. Individual field comparisons can be found in Table \ref{tab:mroz19_comparison}.
    We find that adding in multiples brings the field average $t_E$ into agreement with with \cite{2019ApJS..244...29M}. We also find that the $N_{\rm stars}/\textrm{deg}^2$ is closer to the data and event rate in events/deg$^2$/year is closer and within the errorbars of the data. 
    \label{tab:mroz19_comparison_field_avg}}
    \tablehead{
       \colhead{Field}&  \colhead{Source}  & \colhead{$N_{\mathrm{stars}}/\textrm{deg}^2$} & \colhead{Events/star/year} & \colhead{Events/deg$^2$/\textrm{year}} &
       \colhead{$\langle t_E \rangle_{300}$} & 
       \colhead{Med($(t_E)_{300}$)}
       }
        \startdata
Field Avg & S Runs & 3.40 & 15.86$\pm$1.9 & 66.19$\pm$15.3 & 18.7$\pm$1.2 & 13.8$\pm$0.9 \\ 
 & M Runs & 4.45 & 14.98$\pm$1.8 & 85.05$\pm$20.8 & 25.4$\pm$3.3 & 15.3$\pm$2.1 \\ 
 & Mroz19&6.50 & 15.66$\pm$1.4 & 100.79$\pm$16.1 & 25.6$\pm$1.4 & --  

        \enddata
\end{deluxetable*}

\subsection{Fitting Lightcurves and Parameter Bias}
\label{sec:fitting results}
\begin{figure}
    \centering
    \includegraphics[scale =0.28]{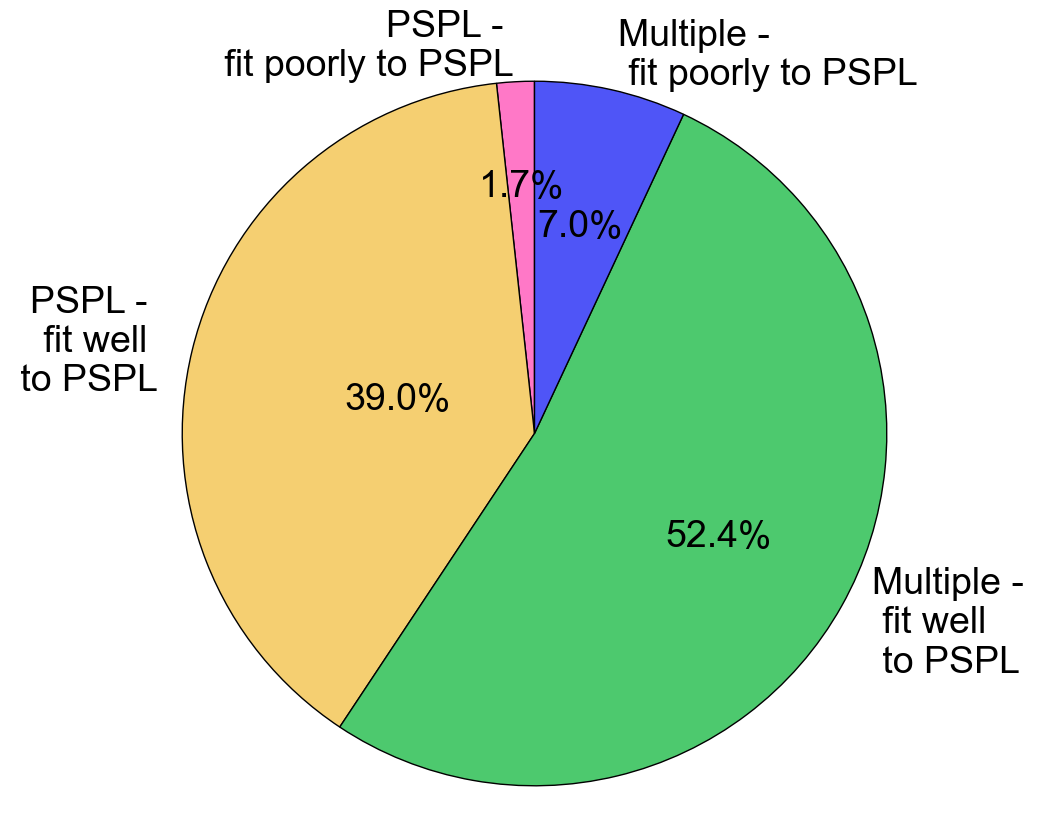}
    \caption{\edit{Simulated events fit by a PSPL model without parallax which first passed Mock OGLE EWS Cuts (see Section \ref{sec:fit lightcurves}) and had significant signal beyond a straight line. 28\% of the fit lightcurves had significant signal beyond a straight line, which are those shown in the pie chart. Of the remaining 1,625 events: 39\% of events were PSPL events that were fit well by a PSPL model (yellow), 52.4\% were multiple event that were fit well by a PSPL model (green), 1.7\% were PSPL events that fit poorly by a PSPL model (pink), and 7.0\% were multiple events that were fit poorly by a PSPL model (blue). Of the PSPL events that were fit poorly by a PSPL model, some had strong parallax signals which were not captured by our non-parallax model and some had poor fits due to the noise we introduced into the data. The majority of multiple events are fit well by PSPL and the majority of events fit well by PSPL are multiple events masquerading as PSPL.}}
    \label{fig:pie chart fit events}
\end{figure}

\edit{In Section \ref{sec:fit lightcurves}, we described fitting the 5,898 lightcurves that passed the Mock-OGLE EWS cuts.}
\edit{28\% of those had significant signal beyond a straight line (see Section \ref{sec:fit lightcurves}). Of those, 91.3\% of events were fit well by PSPL models. As can be seen in Fig.~\ref{fig:pie chart fit events}, 39\% of events were PSPL events that were fit well by a PSPL model, 52.4\% were multiple event that were fit well by a PSPL model, 1.7\% were PSPL events that fit poorly by a PSPL model, and 7.0\% were multiple events that were fit poorly by a PSPL model. Of the PSPL events that were fit poorly by a PSPL model, some had a strong parallax signal that meant they were fit poorly by the model which did not include parallax. Some, though, had a model that tried to fit some of the random scatter that was added and fit the incorrect model. Overall, most of the multiple events were fit well by a PSPL model and most of the microlensing events that were fit well by a PSPL model are in reality not single.}

\begin{figure*}
    \centering
    \includegraphics[scale = 0.5]{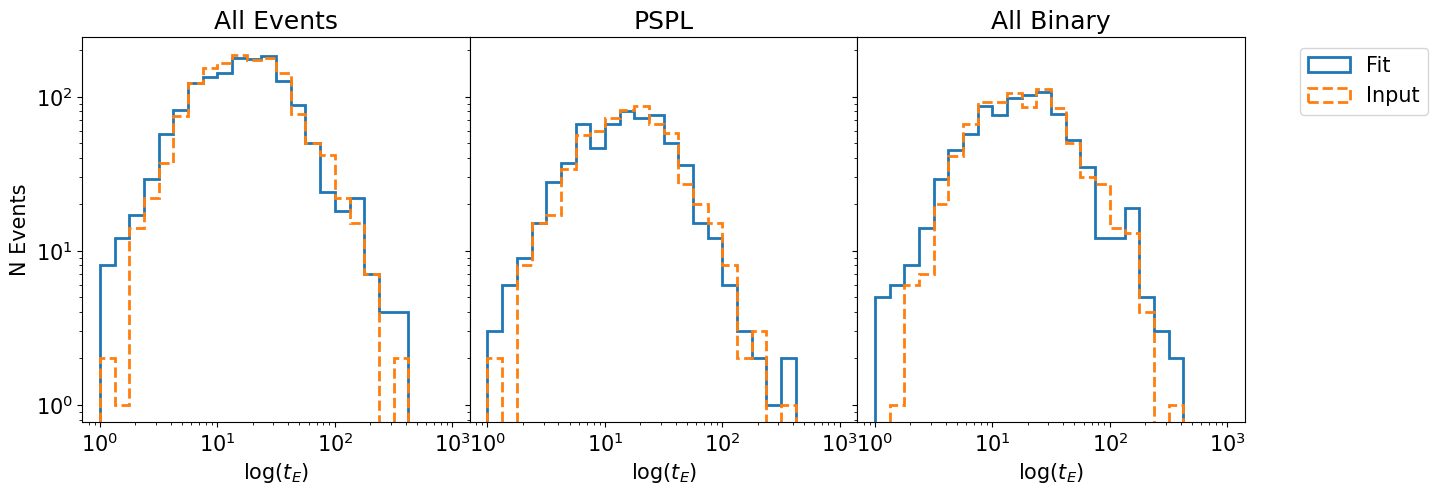}
    \caption{\edit{Events that were fit by a PSPL model without parallax, which first passed Mock OGLE EWS Cuts (see Section \ref{sec:fit lightcurves}). These are events that had a significant signal above a straight line and fit well by a PSPL model. Each panel shows the $t_E$ distributions of the input parameters (dashed orange) and fit parameters (blue). \textit{Left} is all events, \textit{middle} is all PSPL events, and \textit{right} is all multiple events. On a distribution level, there does not appear to be a significant difference in the input and fit parameters. There is some difference in $t_E$ at the low $t_E$ part of the distribution, but we do not explore that in this paper due to the lack of inclusion of brown dwarfs and exoplanets.}}
    \label{fig:fit vs input tE}
\end{figure*}

\edit{In addition, we investigated the parameter bias that may be caused by fitting multiple events with PSPL lightcurves. We found that the overall distributions between the input and fit $t_E$ values (where we use the MAP values for the fit $t_E$ values) did not change significantly (see Fig.~\ref{fig:fit vs input tE}).}
\edit{We also investigated the distributions of bias factors (fit value - input value)/fit error. We performed a Kolmogorov-Smirnov two-sample test between the bias distribution for each parameter and a Gaussian distribution. We found that the $t_E$, $t_0$ and $b_{sff}$ distributions were consistent with a Gaussian with 99.7\% confidence for PSPL, BSPL, PSBL, and BSBL. In both binary and PSPL cases, $u_0$ and $m_{base}$  distributions were inconsistent with a Gaussian to with 99.7\% confidence. In the case of $m_{base}$, this is because the distributions were much narrower than a Gaussian. The $u_0$ values tended to be larger for the input than the fit values.}

\edit{For both PSPL and binaries, there was significant bias on an individual event level, especially in $t_E$ and $u_0$. There are gaps and scatter in the data which may affect how these quantities are fit and degeneracies between them which may cause bias. These problems may be addressed by examining the full posteriors in more detail instead of using the MAP quantities. Since the same issues were faced by both PSPL and multiple events, we accept that the estimates we use for $t_E$, which are based on system mass, are sufficient for the analysis of the paper, and will address general bias in microlensing fitting in future work.}

\subsection{Properties of Binaries Pre- and Post-Microlensing}
\begin{figure}
    \centering
    \includegraphics[width=0.48\textwidth]{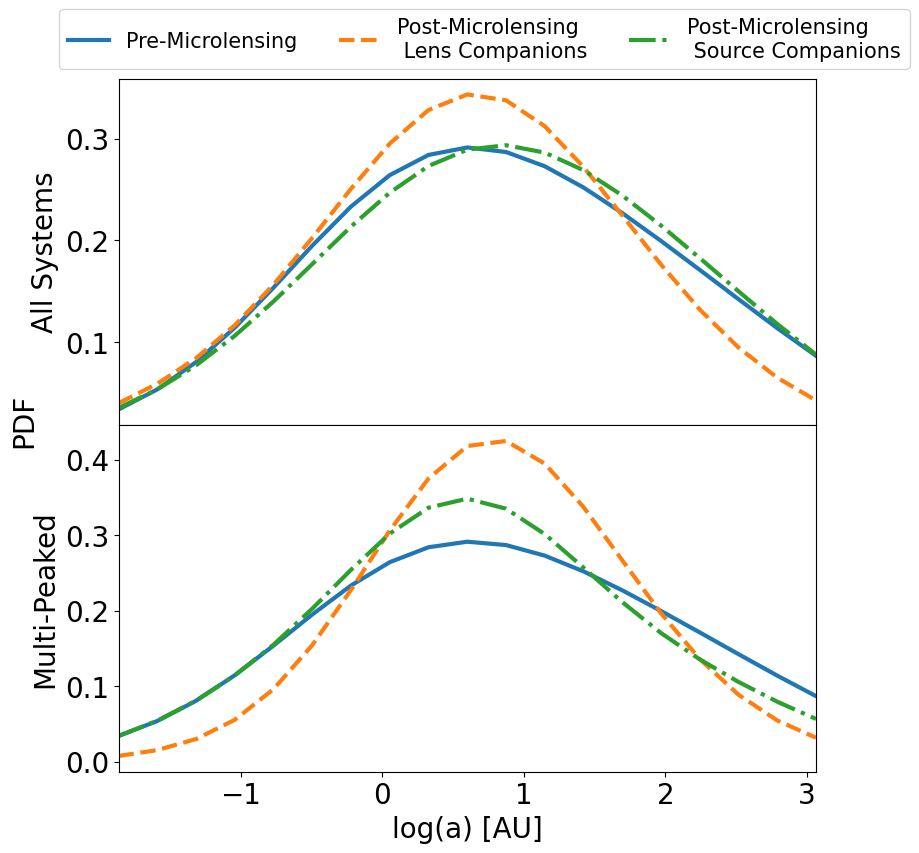}
    \caption{We compare the distribution of semi-major axes of binary systems at the population synthesis stage before microlensing (blue), binaries which are lenses (orange dashed), and binaries which are sources (green dash-dotted). The distributions are smoothed with a Gaussian KDE. The top panel has all systems which shows that microlensing tends to disfavor widely separated binary lenses and favor more widely separated binary sources (these results are the same for single peaked events). The lower panel has only multipeaked systems which shows microlensing tends to disfavor both widely and narrowly separated binary lenses and favor more narrowly separated binary sources in multipeaked events.}
    \label{fig:log_a_pre_post_microlensing}
\end{figure}

\begin{figure}
    \centering
    \includegraphics[width=0.48\textwidth]{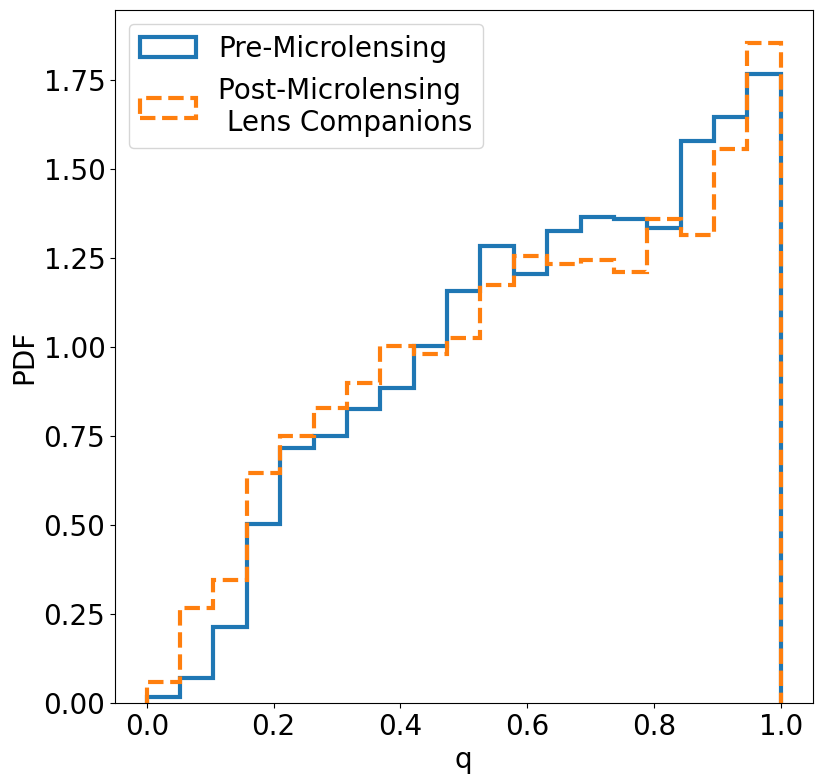}
    \caption{We compare the distribution of q (companion current mass/primary current mass) of binary systems at the population synthesis stage before microlensing (blue) and binaries which are lenses (orange dashed). For any systems where $q > 1$ due to mass loss such that the primary is the less massive object, we take the inverse of q. There seems to be no statistically significant change in q distribution before microlensing and between binary lenses. }
    \label{fig:q_pre_post_microlensing}
\end{figure}

We investigate how input binary parameters, such as $\log(a)$ and $q$, differ from those post-microlensing, since a difference indicates that a certain kind of system is more likely to lens or be lensed. It also informs how, if we are able to statistically measure binary population parameters via microlensing, we will be biased. Our sample here is with Mock OGLE EWS cuts and includes both single and multipeaked multiples. As can be seen in the top panel of Fig.~\ref{fig:log_a_pre_post_microlensing}, we find that binary lenses with a large separation are preferentially not lensed. There seems to be a preferred $\log(a)$ to be lensed between 1-10 AU. In the bottom panel, we see that multipeaked binary lenses are even more biased towards $\log(a)$ values between 1-10 AU, as systems with both small and large separations are less likely to be lensed.
In the top panel of Fig.~\ref{fig:log_a_pre_post_microlensing}, we also see that binary sources are biased towards larger separations.
In the bottom panel of Fig.~\ref{fig:log_a_pre_post_microlensing}, we see that the opposite is true for multipeaked binary sources which are biased towards smaller $\log(a)$.
We also observed a drop-off of multipeaked binary lens events at $\textrm{projected separation} < 0.3\theta_E$ compared to single peaked binary lens events. 
In Fig.~\ref{fig:q_pre_post_microlensing}, we see that the $q$ distributions before microlensing and of the lenses after microlensing are not significantly different. In this case, we are comparing the current mass (as opposed to the ZAMS mass) of the companion/primary, as these are the masses that are observed in the microlensing event.

\subsection{Multiples Effect on Black Hole Astrometric Follow-Up}
\begin{figure*}
    \centering
    \includegraphics[scale=0.55]{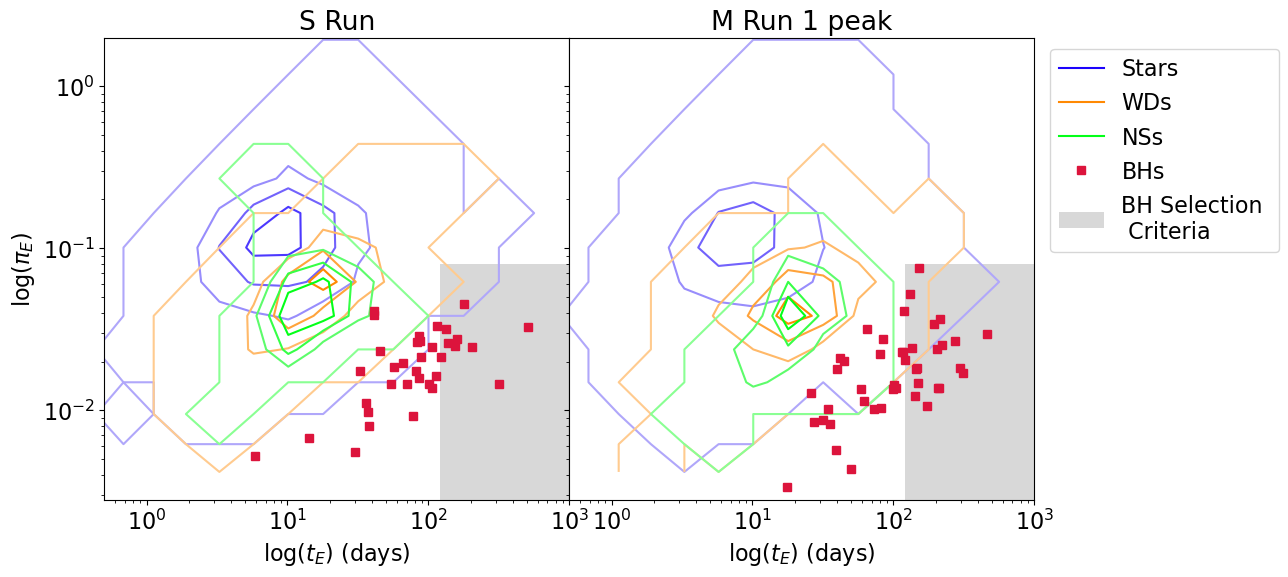}
    \caption{$\pi_E$ vs 
    $t_E$ for S Runs (\edit{left}) and single and single-peaked multiples events for M Runs (\edit{right}). \edit{The blue contours are stars, the orange contours are WDs, the green contours are NSs, and the red squares are BHs. The contours are density contours.} The grey box is the selection criteria which were established in \cite{2020ApJ...889...31L} of $t_E >$ 120 days and $\pi_E <$ 0.08. \edit{The selection criteria was established to capture the area where black holes are mostly isolated from other events, though the bulk of black hole microlensing events are not in this box.} Black holes are \edit{still} mostly isolated in this box in both the M Runs and the S Runs.}
    \label{fig:piE_tE_pspl-like}
\end{figure*}

\begin{figure}
    \centering
    \includegraphics[width=0.48\textwidth]{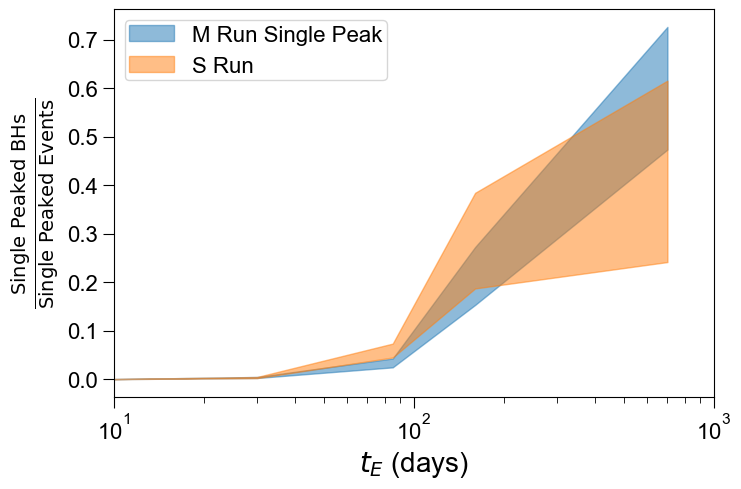}
    \caption{Fraction of events which are black holes as a function of the minimum 
    $t_E$ (Eq. \ref{tE}) per bin. The contours are $1\sigma$ Poisson errors. There is no significant difference in the fraction of black holes at any given $t_E$ between the M Runs and S Runs.}
    \label{fig:BH/all events}
\end{figure}

In \cite{2020ApJ...889...31L}, the limits of $t_E > 120$ days and $\pi_E < 0.08$ were established to optimize for selecting black hole targets for astrometric follow-up. As can be seen in Fig.~\ref{fig:piE_tE_pspl-like}, the black holes continue to be isolated in $\pi_E$-$t_E$ space. Note that in the M Runs all black holes are in multiple star systems since no binary dynamical evolution is currently included (see Sec.~\ref{sec: future improvements}). The contours of M Run non-black hole events overlaps a small fraction of the gray box, which represents the selection criteria. As can be seen in Fig.~\ref{fig:BH/all events}, the fraction of black holes as a function of $t_E$ is indistinguishable between M Runs and S Runs. This indicates that multiples will not affect black hole astrometric follow-up candidate selection.

\section{Discussion}
\label{sec:discussion}

\subsection{Comparison to Other Simulations}
There are a number of other simulations that are used for microlensing population synthesis work. \texttt{genulens} \citep{Koshimoto:2021} developed and optimized a parametric model of the Galaxy based on Gaia \citep{Gaia:2018-disk_kinematics}, OGLE \citep{Nataf:2013}, VVV \citep{VVV-Smith:2018}, and Brava \citep{Brava-Rich:2007, Brava-Kunder:2012} star counts, proper motions, and parallaxes, and the OGLE IV $t_E$ distribution \citep{2017Natur.548..183M, 2019ApJS..244...29M}. They included stellar binary lenses that would be undetectable as determined by their caustic size with physical distributions as described by \cite{2013ARA&A..51..269D} and include compact objects following methods in \texttt{PopSyCLE}. They also included exoplanets and brown dwarfs, which \texttt{PopSyCLE} does not currently include. \texttt{genulens} is useful as a flexible and parametric galactic model and to assess degenerate solutions to microlensing events using Galactic priors. However, \texttt{genulens} has a different purpose than \texttt{PopSyCLE}; it is not designed to generate a set of microlensing events to analyze and, as it uses the OGLE microlensing $t_E$ distribution as part of its training data, it cannot be used to generate $t_E$ distributions  a priori and determine how the $t_E$ distribution changes as the input physics (IMF, IFMR, binary distributions, etc.) change.

\texttt{GULLS} \citep{Penny:2013, Penny:2019}\footnote{In \cite{Penny:2013}, the code was called \texttt{MAB$\mu$LS}, but the name was changed to \texttt{GULLS} to distinguish it from the online tool with the same name described in \cite{Awiphan:2016}.} 
is a microlensing simulation code which is also based on Besan\c con model of the Milky Way. Following the method of \cite{Kerins:2009}, \texttt{GULLS} simulates a population of stars and draws source and lens stars from that population, where a magnitude limit is made on the source stars. Those stars are then randomly matched making a microlensing event population. It compares its numbers of stars and event rates to data and finds \texttt{GULLS} natively produces too few microlensing events and stars, so those values are multiplied by a constant factor 
of 2.33 \citep{Penny:2013} or 2.81 \citep{Penny:2019} 
to correct for that discrepancy. \texttt{GULLS} has the capability to inject their lightcurves into simulated images for survey planning. They also include exoplanets and brown dwarfs, but do not include compact objects or binaries. As can be seen in Fig.~13 of \citep{Penny:2013}, when a $u_0$ and baseline magnitude observational cut are applied, as is similar to \texttt{PopSyCLE} without binaries, the peak of the log($t_E$) distribution is log($t_E) < 20$ days.

\texttt{MaB$\mu$lS} \citep{Awiphan:2016, Specht:2020} is also based on the Besan\c con model of the Milky Way. Like \texttt{GULLS}, they follow the method of \cite{Kerins:2009} by creating catalogs of lenses and sources and finding integrated quantities of optical depth, average $t_E$, average $\mu_{\rm rel}$, and event rate. They include brown dwarfs and high-mass free floating planets and the resulting finite source effects. They find the slope of the IMF of brown dwarfs by fitting to OGLE data \citep{2019ApJS..244...29M}. They also do not include compact objects or binaries. They compare their average quantities as a function of Galactic $\ell$ and $b$ to those of OGLE and mostly find agreement with a slight overestimation of $t_E$ and an underestimation of event rate.

There are a class of simulations that use an analytic model of the Milky Way and make a time-averaged, spatially-averaged mock microlensing survey, pioneered by \cite{HanGould:1995}. In these, one can make predictions for the microlensing $t_E$ distribution \citep{Niikura:2019} and find how the microlensing parameter distributions change by varying underlying Galactic parameters such as the black hole distribution \citep{2020ApJ...905..121A}. In \cite{Toki:2021} they also showed the simulation can be used in a Bayesian study to determine underlying parameters such as mass of the lens from the microlensing lightcurve parameters \citep[\texttt{PopSyCLE} can also be used in this way, see][]{Perkins:2023}. Their simulated $t_E$ distribution can reproduce the $t_E$ distribution from OGLE data \citep{Niikura:2019}.  These simulations include binaries, but in a rudimentary way -- assuming a 40\% binary fraction with all equal mass binaries that are close enough to act as a single lens. 

Similar analytic approaches are used in other papers including those presenting microlensing survey results \citep[e.g.][]{Sumi:2011,  2017Natur.548..183M}. These will often do a comparable calculation, but will maximize the likelihood by optimizing the model to best fit their data. \cite{Wegg:2017} did so specifically to probe the IMF and binary fraction of the Galactic bulge. They put in a distribution of binary parameters, and treated those with small separation as acting as one lens and those with large separations as the primary only. \cite{Wegg:2017} found they could not distinguish between binary fractions when fitting the OGLE $t_E$ data. On the other hand, \cite{2017Natur.548..183M} found their best model had a 40\% binary fraction. In all of these cases, unlike \texttt{PopSyCLE}, they do not simulate individual microlensing events on which we can make observational cuts. They also do not simulate individual objects that have not participated in microlensing which can be used in population statistics work.

\subsection{Future Work}
\label{sec: future improvements}

\texttt{PopSyCLE}'s support for multiplicity, compact objects, and metallicity-dependent IFMRs represents a significant advancement in our ability to simulate microlensing populations. However, there are some limitations and areas we plan to improve in future versions. We currently neglect the orbital motion of binaries, known as xallarap \citep[e.g.][]{Han&Gould:1997, Poindexter:2005}, during the mock survey, which can modify the lightcurves and move some events from single-peaked multiples to multi-peaked multiples. \edit{Currently we also use a mass and period independent mass ratio distribution and a somewhat simplistic eccentricity distribution which we can improve to include mass and separation dependence in future versions.} We also plan to improve the treatment of triple systems. 

In this version of \texttt{PopSyCLE}, we use the system mass to determine observable quantities such as $\pi_E$ and $t_E$; this could lead to a bias in lenses with a massive, distant, and unlensed companion.
For events with triple systems, when simulating their lightcurves to determine observability, we use whichever primary-companion pair leads to the larger $\Delta m$. In reality, triple lens systems should be simulated together to make a single light curve (see Appendix \ref{appendix: triples}).  

Currently, we evolve our stars as singles rather than including binary interactions, so all of our black holes are in binaries. We plan to incorporate a binary evolution code such as \texttt{COSMIC} \citep{Breivik:2020-cosmic} or \texttt{COMPAS}  \citep{Riley:2022-compas} to model the evolution of systems so that disruption of binaries and other effects can be included. Since the distributions of MF and CSF were drawn from real populations of stars \citep{2013ApJ...764..155L}, it is likely that the stellar distributions are relatively accurate. However, there are certainly effects with compact object formation that are likely not accounted for in these numbers such as mergers or natal kicks which could eject objects \citep{Renzo:2019} and lead to a decreased binary fraction or supernovae which could decrease the masses of the resulting objects \citep{Patton2022}.

\edit{As we continue to analyze the affect of binaries on microlensing properties, we will also include additional higher order non-binary effects. In \citep{Golovich:2022}, they fit OGLE microlensing events with parallax and Gaussian process, which accounts for systematic noise, and find that the measured $t_E$ distribution can change from one where the events were fit without these effects. It will be interesting to see how these results are affected by add binarity into the fits.}

The addition of multiples to \texttt{PopSyCLE} will also enable many new theoretical studies such as demographic estimates of stars with wide black hole binaries which can allow us to put astrometric binaries found with Gaia \citep[e.g.][]{El-Badry:2023-BH1, El-Badry:2023-BH2, Chakrabarti:2023} in context; interpreting of surveys searching for black holes via radial velocity \citep[e.g.][]{Clavel:2021}; understanding the possibility of white dwarf self lensing \citep[e.g.][]{Nir:2023};  and quantifying the contamination of exoplanet microlensing with wide binaries \citep[e.g.][]{HanGaudi2008}.

\section{Conclusion}
\label{sec:conclusion}
We found that multiple system sources and lenses make up the majority of observable microlensing events and thus cannot be neglected. Specifically, we found:
\begin{itemize}
    \item Over half of microlensing events likely contain a multiple system as a lens or source, many of which can masquerade as single events.
    \item Adding in multiple systems brings \texttt{PopSyCLE} simulations into agreement with $t_E$ and event rate data.
    \item On average, lens binaries are biased away from large semi-major axes and source binaries are biased towards them. Whereas multipeaked lens binaries are biased towards average semi-major axes and source binaries are biased towards small semi-major axes.
    \item Binaries are unlikely to affect black hole candidate selection.
\end{itemize}

The idea that binaries are important to understanding populations of microlensing events and that they are likely being missed as normal PSPL events is not new. As obvious binary microlensing began to be found, it was conjectured that most microlensing events have binary lenses \citep{DiStefano:2000} and that many are likely easily misclassified as single lenses \citep{1997ApJ...488...55D, 2008ApJ...686..785N}. It was also thought that binary sources may often be missed by being more easily modeled by a blended single source \citep{Dominik:1998}. \cite{GriestHu:1992} estimated that binary sources would be distinguishable from singles sources a minimum of 2-5\% of the time and a maximum of 10-20\% of the time. \cite{MaoPaczynski:1991} estimate $\sim 10$\% of microlensing events are expected have a ``strong" binary signal, but note that binary stars with separations much smaller or larger than their Einstein radius are likely to act like singles. Neglecting multiples could have the impacts outlined here along with potentially leading to overestimates of the optical depth \citep{DiStefano:2000, Han:2005}, but due to the computational challenge of fitting all microlensing events with binary lightcurves, we have yet to systematically do so. As we continue to interpret events from ongoing microlensing surveys and begin to detect microlensing events with new surveys such as the Vera C. Rubin Observatory \citep{2019ApJ...871..205S, 2020ApJ...905..121A, Abrams:2023-Rubin} and the Nancy Grace Roman Telescope \citep{2020AJ....160..123J, Lam:2023-roman}, it is imperative that we incorporate binaries into our understanding and interpretation of their results.

\bigskip
\noindent
We thank Guy Nir, Will Dawson, Scott Perkins, and Peter McGill for helpful conversations. This research used resources of the National Energy Research Scientific Computing Center (NERSC), a U.S. Department of Energy Office of Science User Facility located at Lawrence Berkeley National Laboratory, operated under Contract No. DE-AC02-05CH11231 using NERSC award HEP-ERCAP0023758 \edit{and HEP-ERCAP0026816}. N.S.A., J.R.L., and C. Y. L. acknowledge support from the National Science Foundation under grant No. 1909641 and the Heising-Simons Foundation under grant No. 2022-3542.
C. Y. L. acknowledges support from NASA FINESST grant No.
80NSSC21K2043, the H2H8 foundation, and a Carnegie Fellowship and a Harrison Fellowship.

\software{\edit{\texttt{Numpy} \citep{numpy}, 
\texttt{Matplotlib} \citep{matplotlib}, \texttt{Astropy} \citep{astropy}, \texttt{pandas}  \citep{pandas_v1.5.2, mckinney-proc-scipy-2010}, \texttt{SciPy} \citep{SciPy}, \texttt{pymultinest} \citep{pymultinest}, \texttt{BAGLE}, \texttt{PopSyCLE} \citep{2020ApJ...889...31L}, \texttt{SPISEA} \citep{2020AJ....160..143H}}}

\bibliographystyle{aasjournal}
\bibliography{main}

\appendix
\section{Field Parameters}

In Table \ref{tab:mroz19_comparison}, one can find the field-by-field comparison of S Runs, M Runs, and \cite{2019ApJS..244...29M}. 

\startlongtable
\begin{deluxetable*}{c|c|c|c|c|c|c}
    \centering
    \tablecaption{This table compares the individual field results for singles-only simulations (S Runs) and single-peaked multiples+singles simulations (M Runs) to \cite{2019ApJS..244...29M} values by making the cuts indicated in Table \ref{tab:cuts}. We compare the number of stars/deg$^2$ (where multiple star systems are counted a single blended star), events/star/year, events/deg$^2$/year and the mean $t_E$ between 0.5 and and 300 days. We also compute the median $t_E$ between 0.5 and 300 days. $N_{\rm stars}$ is $10^{-6}$ the value in the simulation. The average between these fields can be found in Table \ref{tab:mroz19_comparison_field_avg}.
    \label{tab:mroz19_comparison}}
    \tablehead{
       \colhead{Field}&  \colhead{Source}  & \colhead{$N_{\mathrm{stars}}/\textrm{deg}^2$} & \colhead{Events/star/year} & \colhead{Events/deg$^2$/\textrm{year}} &
       \colhead{$\langle t_E \rangle_{300}$} & 
       \colhead{Med($(t_E)_{300}$)}
       }
        \startdata
BLG500 & S Runs & 3.37 & 26.14$\pm$2.9 & 88.09$\pm$9.7 & 21.2$\pm$2.0 & 15.5$\pm$2.0 \\ 
 & M Runs & 4.35 & 21.24$\pm$2.3 & 92.39$\pm$10.0 & 20.3$\pm$2.4 & 10.5$\pm$2.4 \\ 
 & Mroz19&4.84 & 23.90$\pm$2.0 & 168.80$\pm$13.7 & 18.8$\pm$1.6 & -- \\ 
BLG501 & S Runs & 8.75 & 32.05$\pm$2.0 & 280.38$\pm$17.4 & 18.4$\pm$1.4 & 12.2$\pm$1.4 \\ 
 & M Runs & 11.43 & 32.52$\pm$1.7 & 371.70$\pm$20.0 & 21.1$\pm$1.5 & 12.1$\pm$1.5 \\ 
 & Mroz19&9.51 & 24.10$\pm$1.4 & 222.90$\pm$12.9 & 20.5$\pm$1.1 & -- \\ 
BLG504 & S Runs & 3.11 & 17.61$\pm$2.5 & 54.79$\pm$7.7 & 17.5$\pm$2.6 & 12.8$\pm$2.6 \\ 
 & M Runs & 4.07 & 18.48$\pm$2.2 & 75.20$\pm$9.0 & 23.1$\pm$2.6 & 16.5$\pm$2.6 \\ 
 & Mroz19&8.47 & 16.90$\pm$1.2 & 134.30$\pm$9.1 & 20.0$\pm$1.2 & -- \\ 
BLG505 & S Runs & 7.06 & 20.99$\pm$1.8 & 148.25$\pm$12.6 & 17.0$\pm$1.4 & 13.1$\pm$1.4 \\ 
 & M Runs & 9.29 & 24.29$\pm$1.7 & 225.60$\pm$15.6 & 18.5$\pm$1.3 & 12.5$\pm$1.3 \\ 
 & Mroz19&13.59 & 22.20$\pm$1.1 & 265.30$\pm$12.8 & 21.8$\pm$1.5 & -- \\ 
BLG506 & S Runs & 3.83 & 21.33$\pm$2.4 & 81.64$\pm$9.4 & 18.9$\pm$2.1 & 12.4$\pm$2.1 \\ 
 & M Runs & 5.14 & 21.93$\pm$2.1 & 112.80$\pm$11.0 & 14.8$\pm$1.3 & 11.0$\pm$1.3 \\ 
 & Mroz19&9.18 & 16.50$\pm$1.1 & 137.40$\pm$8.9 & 28.0$\pm$2.9 & -- \\ 
BLG507 & S Runs & 3.77 & 12.54$\pm$1.9 & 47.27$\pm$7.1 & 14.9$\pm$1.8 & 10.3$\pm$1.8 \\ 
 & M Runs & 5.02 & 13.28$\pm$1.7 & 66.60$\pm$8.5 & 21.8$\pm$3.3 & 14.4$\pm$3.3 \\ 
 & Mroz19&8.01 & 12.30$\pm$0.9 & 87.30$\pm$6.1 & 22.9$\pm$1.4 & -- \\ 
BLG511 & S Runs & 3.64 & 15.04$\pm$2.1 & 54.79$\pm$7.7 & 20.7$\pm$2.5 & 17.2$\pm$2.5 \\ 
 & M Runs & 4.75 & 13.12$\pm$1.7 & 62.31$\pm$8.2 & 22.9$\pm$3.0 & 14.4$\pm$3.0 \\ 
 & Mroz19&9.61 & 13.50$\pm$1.0 & 113.90$\pm$8.1 & 24.5$\pm$2.0 & -- \\ 
BLG512 & S Runs & 5.96 & 17.50$\pm$1.8 & 104.20$\pm$10.6 & 16.5$\pm$1.6 & 10.5$\pm$1.6 \\ 
 & M Runs & 7.79 & 17.24$\pm$1.5 & 134.28$\pm$12.0 & 21.0$\pm$1.9 & 14.5$\pm$1.9 \\ 
 & Mroz19&12.49 & 14.00$\pm$0.9 & 148.50$\pm$9.0 & 24.0$\pm$1.5 & -- \\ 
BLG527 & S Runs & 2.04 & 2.63$\pm$1.2 & 5.37$\pm$2.4 & 12.1$\pm$1.5 & 11.8$\pm$1.5 \\ 
 & M Runs & 2.44 & 2.64$\pm$1.1 & 6.45$\pm$2.6 & 43.6$\pm$10.4 & 41.5$\pm$10.4 \\ 
 & Mroz19&4.54 & 5.50$\pm$0.9 & 21.90$\pm$3.5 & 39.5$\pm$5.0 & -- \\ 
BLG534 & S Runs & 3.26 & 20.77$\pm$2.6 & 67.68$\pm$8.5 & 17.2$\pm$2.3 & 11.2$\pm$2.3 \\ 
 & M Runs & 4.28 & 16.30$\pm$2.0 & 69.83$\pm$8.7 & 16.2$\pm$1.9 & 10.2$\pm$1.9 \\ 
 & Mroz19&6.47 & 17.20$\pm$1.3 & 104.80$\pm$8.1 & 20.8$\pm$1.6 & -- \\ 
BLG535 & S Runs & 2.57 & 10.46$\pm$2.1 & 26.86$\pm$5.4 & 14.7$\pm$2.4 & 9.8$\pm$2.4 \\ 
 & M Runs & 3.39 & 14.88$\pm$2.2 & 50.49$\pm$7.4 & 27.6$\pm$4.8 & 13.2$\pm$4.8 \\ 
 & Mroz19&5.41 & 15.20$\pm$1.1 & 85.20$\pm$6.4 & 25.7$\pm$1.7 & -- \\ 
BLG611 & S Runs & 3.66 & 18.76$\pm$2.3 & 68.75$\pm$8.6 & 17.9$\pm$2.1 & 13.9$\pm$2.1 \\ 
 & M Runs & 4.95 & 17.13$\pm$1.9 & 84.87$\pm$9.5 & 23.2$\pm$3.8 & 12.4$\pm$3.8 \\ 
 & Mroz19&4.95 & 16.20$\pm$1.3 & 70.00$\pm$5.7 & 21.8$\pm$1.6 & -- \\ 
BLG629 & S Runs & 1.49 & 1.44$\pm$1.0 & 2.15$\pm$1.5 & 22.4$\pm$3.0 & 22.4$\pm$3.0 \\ 
 & M Runs & 1.78 & 1.81$\pm$1.0 & 3.22$\pm$1.9 & 49.4$\pm$34.4 & 11.1$\pm$34.4 \\ 
 & Mroz19&3.26 & 3.40$\pm$1.1 & 9.70$\pm$3.1 & 36.7$\pm$7.8 & -- \\ 
BLG645 & S Runs & 0.47 & 2.26$\pm$2.3 & 1.07$\pm$1.1 & 11.3$\pm$0.0 & 11.3$\pm$0.0 \\ 
 & M Runs & 0.57 & 5.63$\pm$3.2 & 3.22$\pm$1.9 & 69.3$\pm$35.5 & 37.5$\pm$35.5 \\ 
 & Mroz19&2.38 & 13.60$\pm$1.9 & 29.20$\pm$4.0 & 34.3$\pm$4.5 & -- \\ 
BLG646 & S Runs & 0.44 & 12.12$\pm$5.4 & 5.37$\pm$2.4 & 34.5$\pm$17.5 & 22.3$\pm$17.5 \\ 
 & M Runs & 0.55 & 5.91$\pm$3.4 & 3.22$\pm$1.9 & 11.4$\pm$3.8 & 8.3$\pm$3.8 \\ 
 & Mroz19&3.64 & 8.30$\pm$1.2 & 26.30$\pm$3.7 & 31.7$\pm$4.1 & -- \\ 
BLG648 & S Runs & 1.24 & 10.36$\pm$3.0 & 12.89$\pm$3.7 & 25.5$\pm$6.2 & 18.3$\pm$6.2 \\ 
 & M Runs & 1.53 & 9.11$\pm$2.5 & 13.97$\pm$3.9 & 19.8$\pm$5.0 & 13.0$\pm$5.0 \\ 
 & Mroz19&2.04 & 18.30$\pm$2.4 & 33.20$\pm$4.4 & 24.0$\pm$2.8 & -- \\ 
BLG652 & S Runs & 2.59 & 21.97$\pm$3.0 & 56.94$\pm$7.8 & 21.0$\pm$2.4 & 14.0$\pm$2.4 \\ 
 & M Runs & 3.47 & 13.92$\pm$2.1 & 48.34$\pm$7.2 & 17.0$\pm$2.8 & 12.6$\pm$2.8 \\ 
 & Mroz19&4.68 & 14.20$\pm$1.6 & 60.00$\pm$6.9 & 24.9$\pm$2.6 & -- \\ 
BLG675 & S Runs & 3.94 & 21.54$\pm$2.4 & 84.87$\pm$9.5 & 14.2$\pm$1.5 & 10.2$\pm$1.5 \\ 
 & M Runs & 5.24 & 20.30$\pm$2.0 & 106.35$\pm$10.7 & 17.1$\pm$1.8 & 10.2$\pm$1.8 \\ 
 & Mroz19&4.03 & 26.50$\pm$2.3 & 95.50$\pm$8.3 & 21.0$\pm$1.8 & --  

        \enddata
\end{deluxetable*}

\section{Microlensing Model Definitions}
\label{sec: model definitions}

In this section, we make the definitions of microlensing configurations explicit. See Table \ref{tab: model definitions} for the standard names and descriptions of these models. We also include how we define $t_E$ in each of these cases and the number of lightcurves simulated in Section \ref{sec: bin lightcurves}. 

In the cases of PSBL, BSPL, and BSBL, we simulate a single lightcurve for the event. However, for triples, we simulate multiple lightcurves and use the one with the largest $\Delta m$. The following are the examples of systems we simulate:
\begin{itemize}
    \item Triple lens and binary source:
    \begin{itemize}
        \item Primary lens + companion lens 1 + primary source + companion source
        \item Primary lens + companion lens 2 + primary source + companion source
    \end{itemize}
    \item Triple lens and triple source:
    \begin{itemize}
        \item Primary lens + companion lens 1 + primary source + companion source 1
        \item Primary lens + companion lens 1 + primary source + companion source 2
        \item Primary lens + companion lens 2 + primary source + companion source 1
        \item Primary lens + companion lens 2 + primary source + companion source 2
    \end{itemize}
\end{itemize}
The parameters for all these lightcurves are stored in the \texttt{lightcurves.fits} table where each bullet point would be an entry in that file. We then choose the lightcurve with the largest $\Delta m$ as the microlensing event. Whether it was used or not is indicated in the lightcurves table.

\begin{deluxetable*}{l|l|l|l|l}
    \centering
    \tablecaption{Model definitions \label{tab: model definitions}}
    \tablehead{\colhead{Name}  & \colhead{Abbreviation} & \colhead{Description} & \colhead{\texttt{$t_E$}} & \colhead{\# lightcurves sim}}
       \startdata
        Point-Source Point-Lens & PSPL & One Source and One Lens & $t_E \propto \sqrt{M_L}$ & 1 \\
        Point-Source Binary-Lens & PSBL & One Source and Two Lenses & $t_E \propto \sqrt{M_{L1} + M_{L2}}$ & 1  \\
        Binary-Source Point-Lens & BSPL & Two Sources and One Lens & $t_E \propto \sqrt{M_L}$ & 1 \\
        Binary-Source Binary-Lens & BSBL & Two Sources and Two Lenses & $t_E \propto \sqrt{M_{L1} + M_{L2}}$ & 1 \\
        Point-Source Triple-Lens & PSTL & One Source and Three Lenses & $t_E \propto \sqrt{M_{L1} + M_{L2} + M_{L3}}$ & 2 \\
        Triple-Source Point-Lens & TSPL & Three Sources and One Lens & $t_E \propto \sqrt{M_L}$ & 2 \\
        Binary-Source Triple-Lens & BSTL & Two Sources and Three Lenses & $t_E \propto \sqrt{M_{L1} + M_{L2} + M_{L3}}$ & 2 \\
        Triple-Source Binary-Lens & TSBL & Three Sources and Two Lenses & $t_E \propto \sqrt{M_{L1} + M_{L2}}$ & 2 \\
        Triple-Source Triple-Lens & TSTL & Three Sources and Three Lenses & $t_E \propto \sqrt{M_{L1} + M_{L2} + M_{L3}}$ & 4
        \enddata
    \tablecomments{These are summaries of the definitions of microlensing configurations included in this paper. $t_E$ is defined as Eq.~\ref{tE} where the $M$ value in $\theta_E$ (Eq.~\ref{thetaE}) is the square root of the total system mass of the lens. The number of lightcurves simulated refers to Section \ref{sec: bin lightcurves} where, since we only have binary models, we simulate multiple lightcurves for triple systems and select the one with the highest $\Delta m$. See the body of Appendix \ref{sec: model definitions} for examples.}
    
\end{deluxetable*}

\section{Triples}
\label{appendix: triples}
10.4\% of all events and 10.1\% of single peaked events have a triple lens or source (see Table \ref{tab: triple breakdown} for a further breakdown). Since we did not simulate triple lens or source lightcurves and instead used the method as described in Appendix \ref{sec: model definitions}, we will verify that there is a significant difference between M Runs and S Runs regardless of the inclusion of triples.

More massive objects are more likely to have a higher number of and more massive companions ($q$ is not dependent on mass, but the same $q$ corresponds to a more massive companion for a massive primary). So, even if we had limited these objects to one companion, they would likely still make up long $t_E$ events. 

We can integrate the total number of single-peaked events with $t_E > 30$ days with and without triples. Roughly 50\% of the difference between M Runs and S Runs is contributed by triples. However, this represents the most extreme conclusion where correctly treated triples would yield multi-peaked or unobservable events instead of single-peaked events. This is unlikely to be the case, in particular, for triple sources, which make up roughly half of all triple systems. 
We conclude that regardless of our treatment of triples, M Runs are a better physical model for the observed $t_E$ distribution.

\begin{deluxetable}{l|l}
    \centering
    \tablecaption{Triple breakdown \label{tab: model definitions}}
    \tablehead{\colhead{Model}  & \colhead{Percentage of all events}}
       \startdata
        PSTL & 2.7\% \\
        TSPL & 4.9\% \\
        BSTL & 1.6\% \\
        TSBL & 0.9\% \\
        TSTL & 0.2\% \\
        \enddata
    \tablecomments{10.4\% of all simulated events have a triple lens or source. This is a breakdown of whether they are triple lenses, sources, or both.}
    \label{tab: triple breakdown}
\end{deluxetable}

\end{document}